\documentclass[aps, preprint, amsfonts,showpacs,longbibliography,citeautoscript,superscriptaddress,]{revtex4-2}
\usepackage[toc,page]{appendix}
\usepackage[british]{babel}
\usepackage[utf8]{inputenc}
\usepackage{amsmath,amssymb}
\usepackage{graphicx}
\usepackage{ae}
\usepackage{esdiff}
\usepackage{braket}
\usepackage{bbm}
\usepackage{simplewick}
\usepackage{float}
\usepackage{braket}
\usepackage{simplewick}
\usepackage{color}
\usepackage{framed}
\usepackage[caption=false]{subfig}
\usepackage[pdftex,colorlinks=true,linkcolor=blue,citecolor=blue,urlcolor=black]{hyperref}
\usepackage{hyperref}
\usepackage{mathtools}
\usepackage{enumitem}
\usepackage{ragged2e}
\usepackage{floatflt}
\usepackage{import}
\usepackage{titlesec}
\usepackage{multirow}
\usepackage{etoolbox}
\usepackage{appendix}
\usepackage{dsfont}
\usepackage{comment}
\usepackage{xcolor}
\usepackage{blindtext}
\usepackage[makeroom]{cancel}
\usepackage{mathtools}
\begin{document}

\newcommand{\norm}[1]{\left\lVert#1\right\rVert}

\def \r{{\boldsymbol{r}}}
\def \lvec{{\boldsymbol{l}}}
\def \x{{\boldsymbol{x}}}
\def \k{{\boldsymbol{k}}}
\def \p{{\boldsymbol{p}}}
\def \q{{\boldsymbol{q}}}
\def \dl{\frac{\partial}{\partial l}}
\def \P{{\boldsymbol{P}}}
\def \K{{\boldsymbol{K}}}
\def \piph{\Pi_\text{ph}}
\def \sign{ \text{sign}}
\def \lamt{\tilde{\lambda}}
\def \intk{{\int_\textbf{k}}}
\def \Ims{\text{Im} [ \Sigma^R(\omega) ]}
\def \Gammabs{\Gamma_{\text{bs}}}
\def \aq{|\q|}
\def \ak{|\k|}
\def \Vt{\tilde V}
\def \omp{\omega^\prime}
\def \om{\omega_m}
\def \on{\omega_n}
\def \omt{\tilde{\omega}}
\def \T{\mathcal{T}}
\def \sgn{\text{sign}}

\definecolor{mgrey}{RGB}{63,63,63}
\definecolor{mred}{RGB}{235,97,51}
\newcommand{\mg}[1]{{\color{mgrey}{#1}}}
\newcommand{\mr}[1]{{\color{mred}{#1}}}

\newcommand{\red}[1]{{\color{red}{#1}}}
\newcommand{\blue}[1]{{\color{blue}{#1}}}
\newcommand{\al}[1]{\begin{align}#1\end{align}}
\newcommand{\beq} {\begin{equation}}
\newcommand{\eeq} {\end{equation}}
\newcommand{\bea} {\begin{eqnarray}}
\newcommand{\eea} {\end{eqnarray}}
\newcommand{\be} {\begin{equation}}
\newcommand{\ee} {\end{equation}}

\def\BigColSep{\setlength{\arraycolsep}{50pt}}

\title{Twists and turns of  superconductivity from a repulsive dynamical interaction}

\author{Dimitri Pimenov}
\author{Andrey V.\ Chubukov}
\affiliation{William I. Fine Theoretical Physics Institute, University of Minnesota, Minneapolis, MN 55455,
USA}

\begin{abstract}
We review recent theoretical progress in understanding spatially uniform $s$-wave superconductivity which arises from a fermion-fermion interaction, which is repulsive on the Matsubara axis, where it is real, but does depend on the transferred frequency. Such a situation holds, e.g., for systems with a screened Coulomb and retarded electron-phonon interaction.  We show that despite repulsion, superconductivity is possible in a certain range of system parameters. However,  at $T=0$ the gap function on the Matsubara axis, $\Delta (\omega_m)$, must pass through zero and change sign at least once.
  These zeros of $\Delta (\omega_m)$  have a topological interpretation in terms of dynamical vortices, and their presence imposes a constraint on the variation of the phase of the gap function along the real frequency axis, which can potentially be extracted from ARPES and other measurements. We discuss how superconductivity vanishes  when the repulsion becomes too strong, and obtain a critical line which terminates at $T=0$ at a quantum-critical point for superconductivity.  We show that the behavior of the gap function near this point is highly non-trivial. In particular, an infinitesimally small $\Delta (\omega_m)$  contains a singular $\delta-$function piece $\omega_m \delta (\omega_m)$.  We argue that near the critical point superconductivity may be a mixed state with even-frequency and odd-frequency gap components.
\end{abstract}

\maketitle

\section{Preface}

Igor Ekhielevich Dzyaloshinskii was one of the greatest physicists of his generation.
   He made seminal contributions to various branches of modern condensed matter physics, including quantum magnetism, superconductivity, and  Fermi-liquid theory.  He is universally recognized as the
   ``father" of the application of the Matsubara-axis formalism to correlated electrons.   In this work we apply his formalism to study an unusual spatially unform superconductivity in systems with repulsive but frequency dependent interaction.  This article is our tribute to a great physicist.

\section{Introduction}

70 years after its discovery, BCS theory \cite{PhysRev.106.162} still forms the basis of our understanding of superconductivity in conventional metals and  in at least some high-$T_c$ materials. At its heart lies Cooper's insight \cite{PhysRev.104.1189} on bound state formation in a Fermi gas: because the density of states near the Fermi level is nearly constant, electrons with momenta $\k$, $-\k$ can form a bound state for arbitrary small \textit{attractive} interaction $V$.

However, while superconductivity is ubiquitous in materials that host itinerant electrons, an
attraction does not appear naturally  because Coulomb repulsion  is typically the largest interaction between electrons.  How to overcome it and get superconductivity? There are two textbook answers.  First, one can look at the momentum dependence of the pairing interaction. In a rotationally invariant systems, one can expand the fully dressed, irreducible interaction $V(\k-\k^{\prime})$ between particles on the Fermi surface in angular momentum components $V_l$ and verify that the analysis of the superconducting instability can be performed individually for each component. For a system to become a superconductor, it is then sufficient for a single partial component $V_l$ to be negative, i.e, attractive.  Kohn and Luttinger have demonstrated that in 3D, the components of the fully dressed $V_l$ with large odd $l$ are necessary attractive, even if all partial components of the bare interaction are repulsive~\cite{PhysRevLett.15.524, RevModPhys.66.129, doi:10.1063/1.4818400, KaganKL}.  The difference comes about because
 in real space the dressed (screened) electron-electron interaction necessary develops Friedel oscillations at large distances, i.e., it occasionally gets over-screened. Partial components  $V_l$ with large $l$ come from large distances, and $V_l$ with odd $l$ predominantly come from distances where the interaction is over-screened, i.e., is attractive.  In 2D the situation is a bit more tricky, but the end result is similar. In a lattice system, the number of orthogonal representations is finite, and there is no generic statement that
 the dressed interaction necessarily has an attractive component.  Yet, in most cases studied in the context of cuprates and other novel superconductors, there exists an attractive component in a pairing channel different from an ordinary $s$-wave.

    Second, $s$-wave superconductivity is also possible if the interaction is repulsive but is retarded and depends on frequency transfer. This is the case when, e.g., the pairing interaction consists of two components: an instantaneous Hubbard repulsion and a smaller retarded attraction mediated by an Einstein phonon with frequency $\Omega_D$ (HEF model). A popular explanation for this behavior is that for large Fermi energy $E_F$, the Hubbard repulsion is logarithmically renormalized down between $E_F$ and  $\Omega_D$, and if this interval is wide enough, the attractive phonon part  prevails at energies below $\Omega_D$~ \cite{tolmachev1958new, Bogoljubov1958,PhysRev.167.331,PhysRev.148.263,PhysRev.125.1263,RevModPhys.62.1027}. This reasoning is a slight oversimplification: both the attractive and repulsive parts are renormalized in the interval between $E_F$ and $\Omega_D$, and the full dressed interaction remains positive, i.e., repulsive.  The more precise argument~\cite{gurevich1962possibility, PhysRev.125.1263, PhysRev.148.263, PhysRevB.28.5100, PhysRevB.94.224515, PhysRevB.96.235107, PhysRevB.100.064513, PhysRevB.98.104505, RevModPhys.62.1027}  is that a non-zero gap function  $\Delta(\Omega_m)$ can develop despite repulsion, but it must have nodes on the Matsubara axis, much like a repulsive momentum-dependent  $V(\k)$ allows a non-zero gap function $\Delta(\k)$ with nodes on the Fermi surface.  The similarity goes even further, as $s$-wave superconducting states $\Delta(\Omega_m)$ with and without nodes on the Matsubara axis are topologically different and in this respect orthogonal, much like in the Kohn-Luttinger scenario $\Delta (\k)$ develops in a spatial channel orthogonal to the ordinary $s$-wave one.

Despite the formal similarities between the momentum- and frequency-dependent repulsive interactions, the frequency-dependent case has some unique characteristics, which are the subject of this article.
These special properties are related to the presence of nodes in $\Delta (\omega_m)$, which, as we will explicitly show below,  are the cores of dynamical vortices.  We show that vortices on the positive Matsubara half-axis either first emerge individually at an infinite Matsubara frequency and then move to a smaller $\omega_m$, or two vortices simultaneously
land on the Matsubara axis at the same $\omega_m$, coming from positive and negative $\omega'$ in the complex plane of frequency, and then split along the Matsubara axis.
 We consider the nodal structure of the gap structure for three models: the continuous HEF model, a toy model with step-like interactions $V(\Omega_m)$, and a model with a $\delta$-functional interaction. The gap in the HEF model contains only a single node, the one in the toy model can contain one or two sign changes, depending on parameters, while the gap in the delta-function model is oscillating.

We then analyze the HEF model in more detail and show that as one increases the strength of the Hubbard term, $\Delta (\omega_m)$ at $T=0$ decreases and finally vanishes at the quantum-critical point (QCP). We show that this is the termination point of $T_c$ as a function of the interaction. We argue that the system behavior near this point is rather intricate as the system needs to keep a balance between the gap amplitude and the location of the nodal (vortex) point, which approaches zero frequency at the QCP.
We show that in the immediate vicinity of the QCP the gap function develops a non-trivial $\delta$-function piece $\omega_m \delta (\omega_m)$ in addition to a regular part. We argue that this
$\delta$-function piece is present in the solution of the linearized gap equation at the QCP.

Finally, we discuss the possible odd-frequency solution for the HEF
model. For a generic case, the conditions for the development of odd-frequency $\Delta_o (\omega_m)$  are much more restrictive than the ones for even-frequency $\Delta_e (\omega_m)$ as for the odd-frequency case there is no Cooper logarithm.  Here, however, odd-frequency pairing becomes a competitor to the even-frequency one because Hubbard repulsion cancels out from the gap equation for $\Delta_o (\omega_m)$.  We argue that the most likely outcome is a mixed superconducting state with both $\Delta_e$ and $\Delta_o$ present.  We show that the relative phase between the two components is $\pm \pi/2$, where the sign is chosen spontaneously. Such a state breaks time-reversal symmetry.

The structure of the remainder of this article is as follows:
 In Sec.\ \ref{vortexsec} we review general analytic properties of a dynamical gap function on the Matsubara axis and consider the nodal structure of $\Delta (\omega_m)$ for a toy model with a step-like repulsive interaction, as well as a model with a delta-function interaction.
 In Sec.\ \ref{evenfreqseq}, we analyze the gap function in the  HEF model and discuss the critical properties near the phase transition from the superconductor to the normal state. We show that the solution of the linearized gap equation contains a $\delta-$functional term. In Sec.\ \ref{OFsec},  we consider the odd-frequency solution $\Delta_o (\omega_m)$  for the HEF model and its interplay with the even-frequency solution.  In Sec. \ref{concsec} we present our conclusions and a list of open questions.

The analysis of the gap function  in this article   is performed  within   the
  Eliashberg theory \cite{eliashberg1960interactions, MARSIGLIO2020168102}.
   An alternative is to start from the Hubbard-Holstein lattice model and solve for superconductivity numerically. See Refs.\ \cite{Bauer_2012, PhysRevB.87.054507, MARSIGLIO199521} and in particular \cite{marsiglio2022impact}  for
   the discussion of this approach and for references to earlier papers.

\section{General properties of dynamical gap functions at large repulsion}
\label{vortexsec}

Our starting point for the analysis of the dynamical gap function are the Eliashberg equations for the
pairing vertex $\Phi(\omega)$ and the self-energy $\Sigma(\omega)$. The Eliashberg treatment, which neglects vertex corrections, is justified  at $T=0$ when $E_F \gg \Omega_D$, as first observed by Migdal \cite{migdal1958interaction}.

In the following, we use frequencies $\omega, \Omega$ without Matsubara index when analyzing  zero-temperature properties, and $\omega_m, \Omega_m$ in the finite temperature case. Neglecting the self-energy at first, the pairing vertex is equivalent to the gap function $\Delta(\omega)$, and the gap equation for purely frequency-dependent interactions reads
\begin{align}
\label{maineq}
\Delta(\omega) = - \frac{\rho}{2} \int_{-\Lambda}^\Lambda d\omp \frac{V(\omega - \omega')}{\sqrt{(\omp)^2 + |\Delta(\omp)|^2}} \times \Delta(\omega') \ .
\end{align}
Here, $\Lambda$ is a UV cutoff of order $E_F$, and $\rho$ the density of states. To understand the structure of  $\Delta(\omega)$, it is convenient to analyze the linearized gap equation at $T = T_c -0$:
\begin{align}
\label{lineareqfirst}
\Delta(\omega) = - \frac{\rho}{2} \int_{-\Lambda}^\Lambda d\omp \theta(|\omp| - T) \frac{V(\omega - \omega')}{|\omp|} \times \Delta(\omega') \ .
\end{align}
The potential infrared  singularity at $\omp = 0$ is cut off by a finite temperature $T$. The gap function $\Delta(\omega)$ is defined to up to a $U(1)$ phase and for convenience we choose it to be real. The gap equation (\ref{lineareqfirst}) allows  two types of solutions --  gap functions which are even in frequency (EF),
      $\Delta (\omega) = \Delta (- \omega)$, and odd in frequency (OF),  $\Delta (\omega) = - \Delta (-\omega)$. The linearized equation decouples between the two components:
\begin{align}
\label{evenoddlin}
\notag
&\Delta_{e/o}(\omega) = - \frac{\rho}{2} \int_{-\Lambda}^\Lambda d\omega^\prime \theta(|\omp| - T) \frac{V_{e/o}(\omega, \omp)} {|\omp|} \times \Delta_{e/o}(\omp) = -\rho \int_T^\Lambda  \frac{V_{e/o}(\omega, \omp)} {|\omp|} \times \Delta_{e/o}(\omp) \\ &
V_{e/o}(\omega, \omp) = V(\omega - \omp)  \pm V(\omega + \omp)
\end{align}
For the remainder of this section and in the next two sections we consider EF pairing and define $\Delta (\omega) = \Delta_e (\omega)$. We analyze OF pairing in Sec.\ \ref{OFsec}.

The goal of this article is to analyze $s$-wave superconductivity coming out of a repulsive interaction  $V(\Omega) > 0$. In this case, a conventional sign-preserving solution $\Delta (\omega)$ cannot develop as for such a gap function the left and the right hand side of Eq.\ \eqref{maineq} have different signs.  One can, however, search for a solution which changes sign along the Matsubara axis. For such a solution there is at least a possibility to change the overall minus sign on the r.h.s. of the gap equation by choosing opposite signs of $\Delta (\omega)$ and $\Delta (\omega + \Omega)$ for relevant bosonic frequencies $\Omega$.

What kind of repulsive interactions allow a sign-changing solution of the gap equation? To get some insight, we first consider  a simplified model with a step-like interaction (Rietschel-Scham model) \cite{PhysRevB.28.5100, PhysRevB.100.064513}:
\begin{align}
V(\Omega) = \begin{cases} &\frac{1}{\rho}  \chi_1 \quad  |\Omega| < \Omega_1 \\ &
\frac{1}{\rho}  \chi_2 \quad   \Omega_1 < |\Omega| < \Lambda \end{cases}  , \quad \quad \chi_1, \chi_2 > 0
\end{align}
We search for a solution to the linear gap equation \eqref{evenoddlin} in the form
\begin{align}
\Delta(\omega) = \begin{cases}  & \Delta_1 \quad 0 < \omega < \Omega_1
\\ & \Delta_2 \quad \Omega_1 < \omega < \Lambda
\end{cases}
\label{n_3}
\end{align}
We assume that $T \ll \Omega_1$ , $\Omega_1 \ll \Lambda$. Under these assumptions  we can solve the gap equation to the leading logarithmic order in  $L_1 \equiv \log(\Omega_1/T)$ and
$L_2 \equiv \log(\Lambda/\Omega_1)$. Substituting (\ref{n_3}) into (\ref{evenoddlin}) for EF pairing, we obtain a system of equations
\begin{align}
\label{eqsystem1}
&\Delta_1 = -\chi_1 \Delta_1 L_1 - \chi_2 \Delta_2 L_2  \\
&\Delta_2 = - \chi_2 \Delta_1 L_1 - \chi_2 \Delta_2 L_2  \ .
\end{align}
This set has a nonzero solution at $T=T_c$ when
\begin{align}
L_1 = \log(\Omega_1/T_c) =  \frac{1}{(\chi_2 - \chi_1) - \frac{\chi_2}{1+ \chi_2 L_2} } {\equiv \frac{1}{\chi_\text{eff}}}\ .
\label{L1eq}
\end{align}
For this $L_1$ the ratio $\Delta_2/\Delta_1$ is
\begin{align}
\frac{\Delta_2}{\Delta_1}  = - \frac{L_1 \chi_2}{1+L_2 \chi_2} \ .
\end{align}
Therefore, the gap function changes sign between small and large frequencies, as expected.

\begin{figure}
\centering
\includegraphics[width=\textwidth]{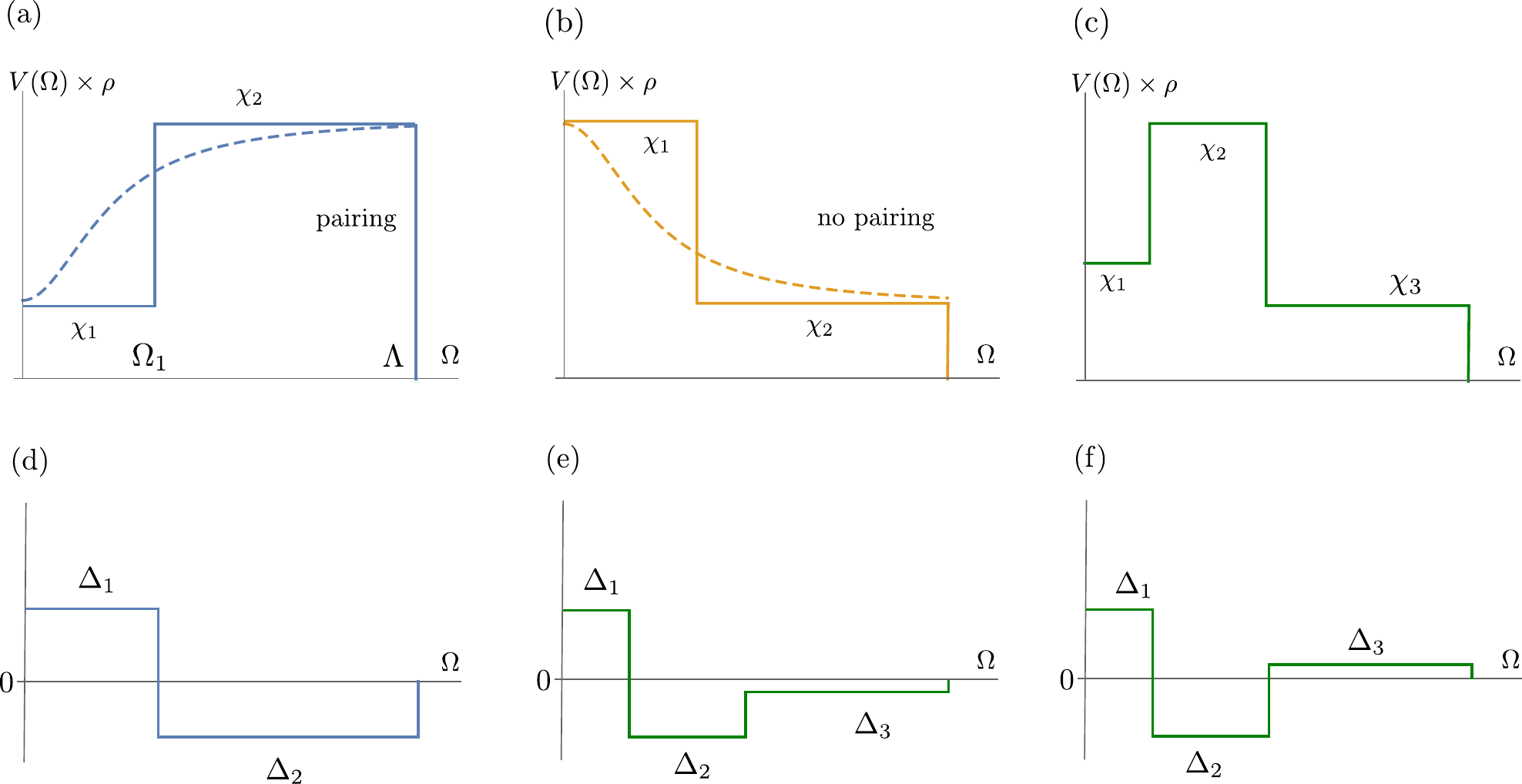}
\caption{Upper panel: (a) Model interaction with pairing. The dashed line corresponds to a more physical electron-phonon type interaction (HEF model) as discussed in Sec.\ \eqref{evenfreqseq}. (b) Model interaction without pairing. The dashed line is a would-be ``repulsive electron-phonon" interaction.
Lower panel: (d) Resulting gap function in the model (a). (e) gap function in the model (c) for $\chi_3 > 0$. (d) gap function in the model (c) for  $\chi_3 < 0$.  }
\label{RSfig}
\end{figure}

We see that a non-zero solution is possible if $\chi_1 < \chi_2^2 L_2/(\chi_2 L_2 + 1) < \chi_2$. This implies that the interaction has to be less repulsive at lower frequencies. This condition is satisfied for a model with a frequency independent Hubbard repulsion and electron-phonon attraction as the latter is larger at small frequencies (Fig.\ \ref{RSfig}(a)). By contrast, an interaction which is less repulsive at large frequencies,  as sketched in Fig.\ \ref{RSfig}(b),
  does not lead to pairing.
  Next, the  r.h.s.\  of (\ref{L1eq}) is the difference between the effective attractive coupling $\chi_2 - \chi_1$ and the
  renormalized repulsive one, {$\frac{\chi_2}{1+\chi_2 L_2}$}. The {bare} repulsive coupling $\chi_2$ is larger than $\chi_1$, but it is
    reduced by $1/(1+ \log(\Lambda/\Omega_1) \chi_2)$. This is known as
Tolmachev-McMillan logarithm \cite{tolmachev1958new, PhysRev.167.331} or Anderson-Morel pseudopotential $\mu^\star$ \cite{PhysRev.125.1263}.

It is tempting to interpret the result as if the repulsive part of the interaction renormalizes down, and at small $\Omega$ the pairing interaction is attractive. This is a bit of an oversimplification as $V(\Omega)$ is repulsive at all frequencies.
 The effective attraction emerges for an effective low-energy model, in which fermions with frequencies $\omega > \Omega_1$ are integrated out. {To see this more clearly, we note that the equation for $\Delta_1$, Eq.\ \eqref{eqsystem1}, can be rewritten as
\begin{align}
\label{RSDelta}
\Delta_1 = \chi_{\text{eff}} L_1 \Delta_1 ,
\end{align}
with $\chi_\text{eff}$ as in Eq.\ \eqref{L1eq}. Therefore, the low-energy behavior  $(\omega < \Omega_1)$ is described by a BCS-like equation with coupling $\chi_{\text{eff}}$; when this coupling becomes attractive, superconductivity becomes possible. }
This effective description is correct, but it sweeps under the rug 
 the information that the full $\Delta (\omega)$ is sign-changing.
   We will see below that the sign change is crucial for the understanding of the disappearance of superconductivity once the repulsion becomes too strong.

   Since a  ``step up" potential with $\chi_1 < \chi_2$ leads to pairing, while a ``step down" potential with $\chi_2 < \chi_1$  does not, it is interesting to consider a combination of these two, i.e., a ``step up-step down" potential, see Fig.\ \ref{RSfig}(c). If we assume that the cutoff for the third region is $\tilde\Lambda \gg \Lambda$, we can solve for the gap function using the ansatz
\begin{align}
\Delta(\omega) = \begin{cases}  & \Delta_1 \quad 0 < \omega < \Omega_1
\\ & \Delta_2 \quad \Omega_1 < \omega < \Lambda
\\ & \Delta_3 \quad \Lambda < \omega < \tilde \Lambda
\end{cases}
\end{align}
Substituting into (\ref{evenoddlin}) for EF pairing, we obtain a set of three coupled equations
\begin{align}
&\Delta_1 = -L_1 \Delta_1 \chi_1 - L_2 \Delta_2 \chi_2 - L_3 \Delta_3 \chi_3 \\
&\Delta_2 = -L_1 \Delta_1 \chi_2 - L_2 \Delta_2 \chi_2 - L_3 \Delta_3 \chi_3 \\
& \Delta_3 = - L_1 \Delta_1 \chi_3 - L_2 \Delta_2 \chi_3 - L_3 \Delta_3 \chi_3
\end{align}
Assuming $\chi_3< \chi_2$, we find that superconductivity always appears at small enough $\chi_1$, and
a generic solution only has a single sign change:  $-\sign(\Delta_1) =  \sign(\Delta_2) = \sign(\Delta_3)$.
If $\chi_1 \ll \chi_2$,  $\Delta_3$ becomes small, and  for negative $\chi_3$ (an attraction at large frequencies), the gap function changes sign twice. We sketch this in Fig.\ \ref{RSfig}.

\begin{figure}
\centering
\includegraphics[width=.5\textwidth]{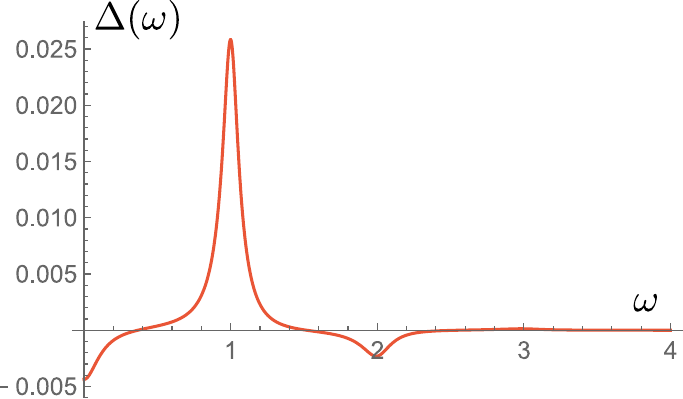}
\caption{Numerical solution of the non-linear gap equation, for a repulsive interaction which is a regularized $\delta$-function with $\omega^\star = 1$.  }
\label{deltafig}
\end{figure}

The appearance of multiple sign changes is generally expected for a repulsive  interaction peaked at some finite frequency. The limiting case of such $V(\Omega)$ is a $\delta-$function:
\begin{align}
V(\Omega) = \lambda \delta(\Omega - \omega^\star) \ .
\end{align}
Because $\Delta (\omega)$ and $\Delta (\omega + \omega^*)$ must have different signs for all $\omega$,
the gap function necessary oscillates with period $\omega^\star$. E.g., $\Delta (\omega) = \cos(\pi \omega/\omega^\star)f(\omega/\omega^\star)$, where $f(x)$ is a sign-preserving function. A numerical solution of the non-linear gap equation for a repulsive interaction of $\delta$-function type with $\omega^\star = 1$ is shown in Fig.\ \ref{deltafig}. Note that the gap equation relates $\Delta(0) \simeq - \lambda \Delta(\omega^\star)$. Therefore $\Delta(0) \ll \Delta(\omega^\star)$ at weak coupling, as seen in the Figure.

The zeros of the Matsubara gap function $\Delta(\omega)$ carry a special meaning, as they are the centers of dynamical vortices. This can be seen by  analytically continuing  $\Delta(\omega)$ to a neighborhood of the nodal point in the frequency upper half plane,  and by studying the phase variation of the complex $\Delta(z)$ along a small circle centered at the nodal point.  The nodes on the Matsubara axis also affect the behavior of the retarded $\Delta (z)$ infinitesimally close to real axis, at  $z = \omega' + i\delta$. The existence of a node at $z = i\omega_0$ implies
  that the phase $\eta(\omega')$ of $\Delta^R(\omega')$ winds by $2\pi$ between large positive and large negative frequencies \cite{PhysRevB.103.024522, PhysRevB.104.L140501}.
    This phase variation is additional to a bare variation in the absence of poles. Its existence   follows from the general
    ``argument principle" for an analytic function in the upper half-plane of frequency
    \cite{brown2009complex}.

Because the phase of the gap changes by $2\pi$, the real and the imaginary part of $\Delta^R(\omega')$ must have nodes as well, see Fig.\ \ref{phasewinding}. These nodes  can potentially be extracted from ARPES~\cite{Damascelli_2004} or other spectroscopic technique, which are sensitive to the complex gap function on the real axis, like tunneling I-V measurements \cite{Marsiglio:2008aa}.
We note in this regard that there cannot be any anti-vortices as  these would correspond to poles of the gap function by the argument principle, which would be inconsistent with analyticity.

\begin{figure}
\centering
\includegraphics[width=\textwidth]{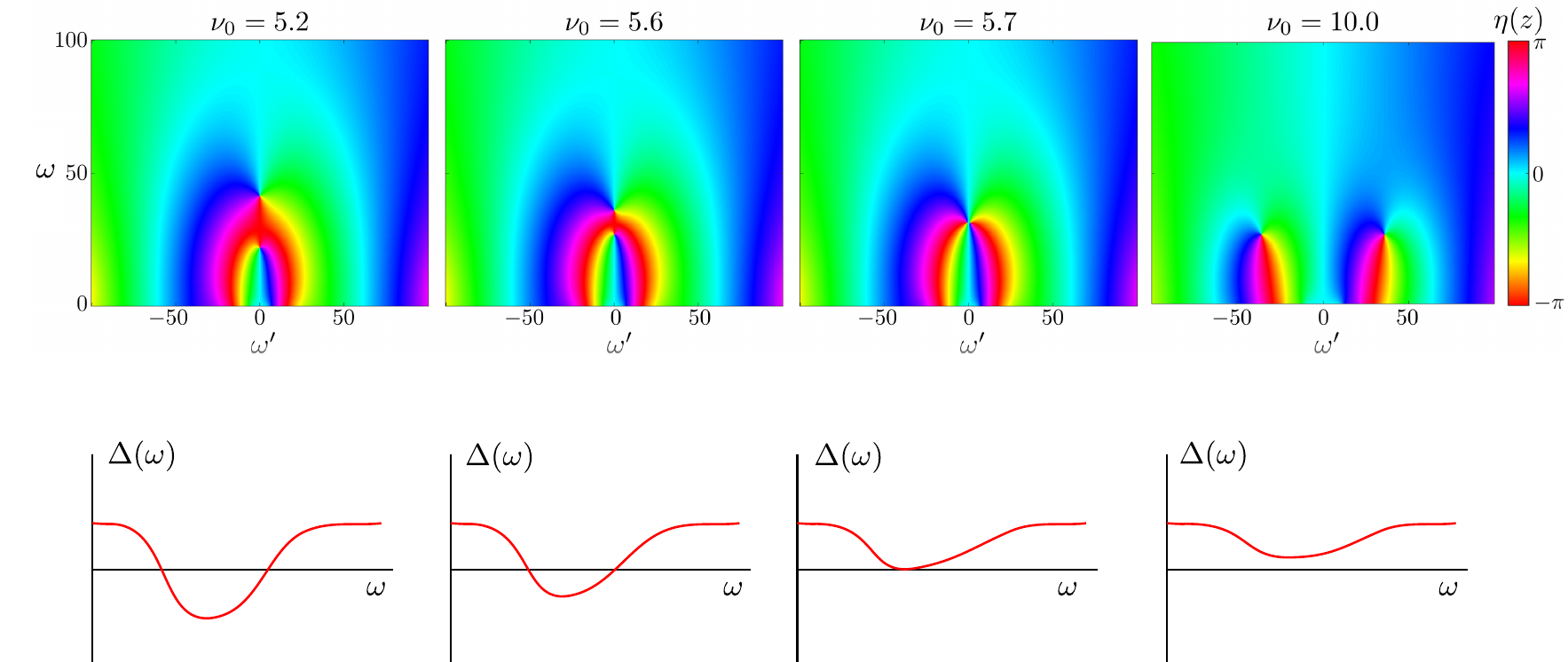}
\caption{Complex gap function $\Delta(z = \omega' + i\omega)$ for an electron-phonon type interaction $V$ with both attractive and repulsive regions, adapted from \cite{PhysRevB.104.L140501}. Upper panel: phase of the gap function as an interaction parameter $\nu_0$ is tuned. Lower panel: \textit{Sketch} of the corresponding gap function on the Matsubara axis. }
\label{Morten1}
\end{figure}

\begin{figure}
\centering
\includegraphics[width=.8\textwidth]{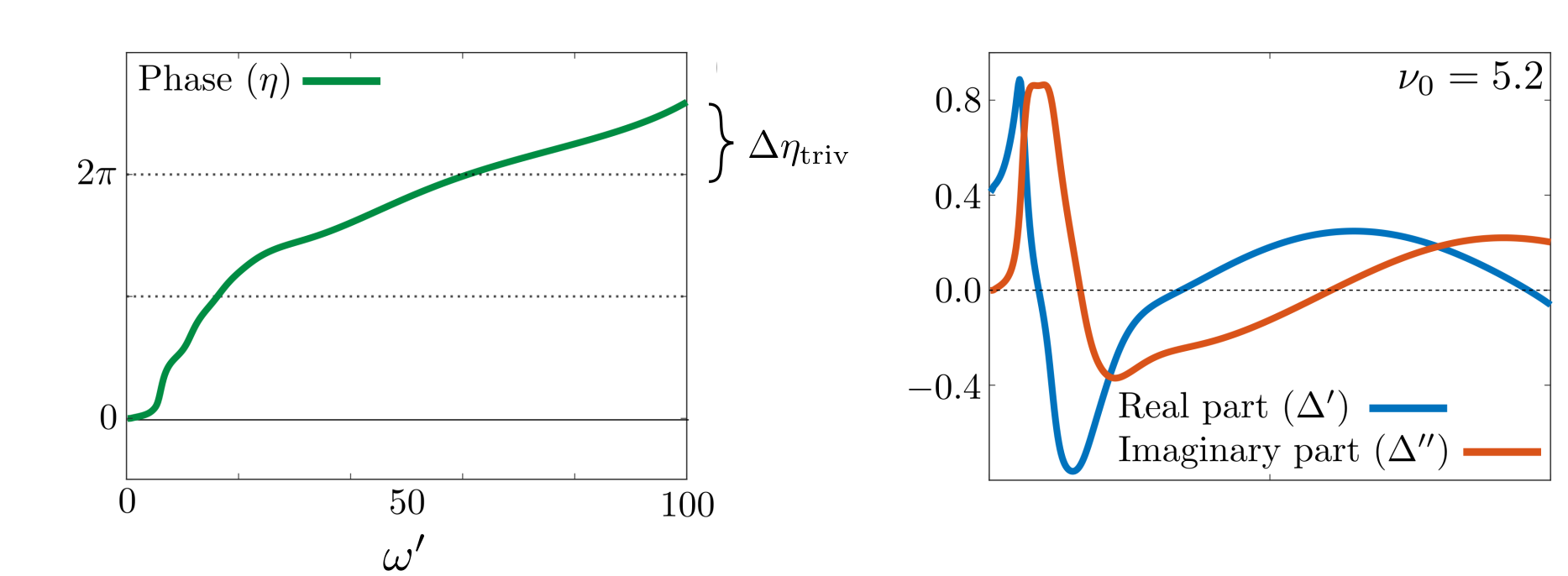}
\caption{Phasewinding $\eta(\omega')$ on the real half-axis of the gap function corresponding to the first column of Fig.\ \ref{Morten1}. There are two vortices, which lead to a phase winding of $4\pi$ on the full real axis, respectively $2\pi$ on the half-axis. The part $\Delta \eta_\text{triv}$ indicates the bare phase winding in the absence of vortices. Fig.\ adapted from \cite{PhysRevB.104.L140501}. }
\label{phasewinding}
\end{figure}

Because a vortex cannot be annihilated by an antivortex, the total number of vortices is a topological invariant: it cannot be changed upon smooth local deformation of the gap function. A single vortex can appear or disappear only at the upper boundary, which in our case is $|\omega| = \Lambda$, see Sec.\ \ref{evenfreqseq}.  Pairs of vortices, however, can move to the upper half-plane from the lower one. One vortex appears at real $\omega'$, another at $-\omega'$. Once in the upper half-plane, vortices can move and can merge on the Matsubara axis.  After they merge, vortices split along the Matsubara axis, creating two new nodal points. We illustrate this in Fig.\ \ref{Morten1} adapted from Ref.\  \cite{PhysRevB.104.L140501}.  Still, every vortex inside the upper half-plane, no matter at which $z$ it is, gives rise to $2\pi$ variation of the  phase of the gap, $\eta (\omega')$ between large negative and large positive (real) $\omega'$.

\section{HEF model }
\label{evenfreqseq}

Let us now focus on a specific interaction $V (\Omega)$, which is approximate but analytically tractable
 and provides a realistic description of electron interactions in a metal \cite{PhysRev.125.1263, PhysRevB.28.5100,PhysRevB.100.064513,PhysRevB.94.224515,PhysRevB.96.235107, PhysRevB.98.104505, PhysRevB.105.064518, doi:10.1143/JPSJ.80.044711, pimenov2021quantum, pimenov2022odd}:
\begin{align}
\label{Vdef}
 V(\Omega) = \frac{2}{\rho} \times \chi(\Omega), \quad \chi(\Omega) =  \lambda \left(f - \frac{\Omega_D^2}{\Omega_D^2 +\Omega^2} \right) .
\end{align}
Here, $\chi$ is a dimensionless interaction, $\lambda$ is the coupling strength, and $f$ is a Hubbard-like repulsion, mimicking a screened Coulomb interaction between electrons. The term with $\Omega_D$
 represents an attractive interaction mediated by an Einstein optical phonon. In the following, we measure all energies in units of $\Omega_D$ and set $\Omega_D \equiv 1$. {We assume a UV cutoff $\Lambda \gg 1$.}

\begin{figure}
\centering
\includegraphics[width=\columnwidth]{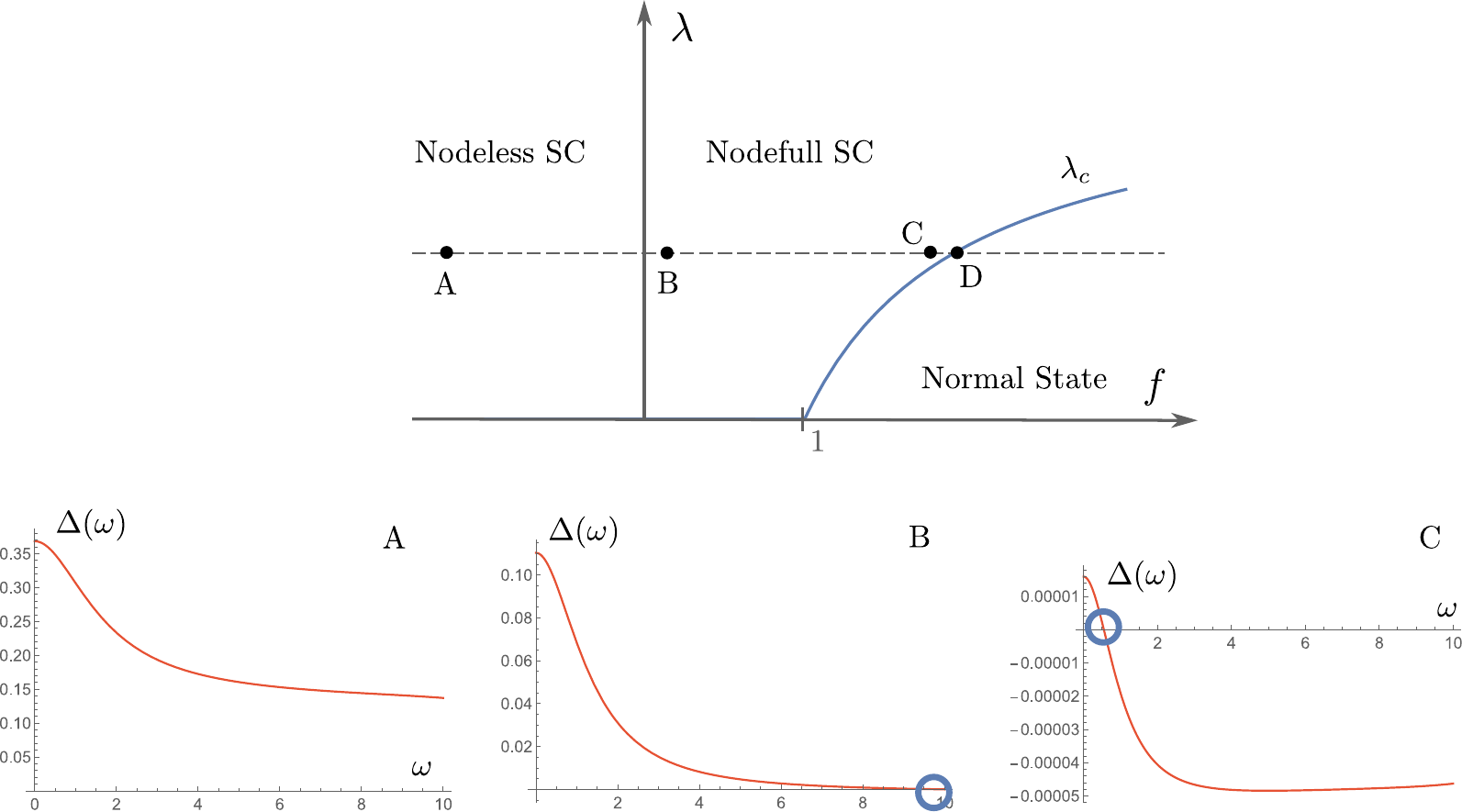}
\caption{Upper panel: $T = 0$ phase diagram of the model \eqref{Vdef}. Numerical solutions for the gap functions at the points A, B, C are shown in the lower panel, while the point $D$ is analyzed in Fig.\ \ref{gammaplot}. Blue circles mark the node position. Note the different units on the vertical axes.  }
\label{phasediag1}
\end{figure}

 The solution of the gap equation at $T = 0$ yields an even-frequency  gap function  $\Delta(\omega) = \Delta(-\omega)$.
   The structure of $\Delta (\omega)$ and the value of $T_c$  strongly depend on the repulsion strength $f$. We can track the evolution of $T_c$ and of the form of $\Delta (\omega)$ by fixing $\Lambda, \lambda$ and gradually increasing $f$, see Fig.\ \ref{phasediag1} \cite{pimenov2021quantum, pimenov2022odd}.

For negative $f$, the interaction is purely attractive, $T_c \propto e^{-a/\lambda}$, where $a = O(1)$, and the gap function is nodeless. For  $0<f<1$, the interaction remains attractive at small $\Omega$, but becomes repulsive at larger $\Omega$. The analysis of the gap equation shows that at  $f = 0^+$, a single node appears  in $\Delta (\omega)$ at $\omega_0 = \infty$. As the positive $f$ increases,  the nodal point moves down to a finite frequency, and its
position scales as  $\frac{1}{\sqrt{f}}$ at small $f$.  The critical temperature still scales as $e^{-a/\lambda}$ and remains finite for any value of $\lambda$, but the prefactor $a$ increases with increasing $f$ and diverges at $f =1$.  At this $f$,  $T_c$ is still finite at arbitrary small $\lambda$, but scales as $e^{-{\bar a}/\lambda^2}$  where ${\bar a} =O(1)$. Consequently, the location of the node of  $\Delta (\omega)$  remains at a finite $\omega_0$ for any non-zero $\lambda$.

   At  $f \geq1$, the interaction becomes purely repulsive.  By continuity, superconductivty persists, but it now requires         $\lambda$ to exceed a  critical  $\lambda_c$. For large $\log(\Lambda)$,
\begin{align}
\label{lambdac}
\lambda_c = \frac{f-1}{2f \log(\Lambda) } \ .
\end{align}

By inverting \eqref{lambdac}, we can obtain an expression for the critical repulsion at a fixed $\lambda$:
\begin{align}
\label{fcvalue}
f_c = \frac{1}{1- 2\lambda \log(\Lambda)} \ .
\end{align}
Superconductivity develops for $f <f_c$.  Note that  $f_c$ diverges at $\lambda = 1/(2 \log(\Lambda))$.
 At larger $\lambda$, $f_c$ is infinite, i.e., for any $f$  the system manages to adjust the position of the node in $\Delta (\omega)$ to keep a superconducting ground state.  This last result may be a peculiarity of Eliashberg theory.
 The analysis of the Hubbard-Holstein model~\cite{marsiglio2022impact} suggests that there exists a
   a maximal upper value for the repulsion, above which superconductivity does not develop.

Below we focus on 
 the parameter range 
 where $f_c$ is finite.  As $f$ approaches $f_c$,  the overall magnitude of $\Delta (\omega)$ decreases and simultaneously the nodal point of $\Delta (\omega)$ must move towards $\omega =0$.  It cannot remain at a finite frequency, because if it was there, one would not be able to solve the gap equation at $T=0$ and $f =f_c-0$  due to an un-regularized Cooper logarithm. The detailed analysis of the behavior of $\Delta (\omega)$ at $T=0$ near the phase transition at $f_c$ shows~\cite{pimenov2021quantum} a non-trivial correlation between the overall magnitude of the gap and the position of the node at $\omega = \omega_0$:
  the node position  $\omega_0$ vanishes as a power law, while the
   magnitude  of the gap function at zero frequency, $\Delta(0)$,  vanishes exponentially:
\begin{align}
\omega_0 \sim \sqrt{f_c - f},  \quad  \quad \Delta(0) \sim
\exp(-1/(f_c -f)) \ .
\label{n1ref}
\end{align}
 The gap function at $\omega >\omega_0$ also vanishes exponentially, but is parametrically larger than $\omega_0$:
 \begin{align}
 \Delta (\omega \geq \omega_0) \sim \Delta (0)/(f_c-f) \ .
 \label{n2ref}
 \end{align}
This inter-locked behavior of $\omega_0$, $\Delta (0)$, and $\Delta (\omega \geq \omega_0)$
  is the only way to solve the gap equation \eqref{maineq}:  like we just said, if   $\omega_0$ remained finite at $f \to f_c$,  the right hand side of the gap equation would   contain an incurable infrared logarithmic singularity, hence there would be no solution of the gap equation.
    On the other hand, if we just set $\omega_0 =0$, there would be no frequency range where the gap changes sign, and
    again there  would be no solution of the gap equation as one cannot get a      sign-preserving
     $\Delta (\omega)$ for a  repulsive interaction.

 The parametric relation $\Delta(0) \ll \omega(0)$  implies that the gap function looks almost constant for small frequencies. As a result, some quantities show BCS-like behavior close to the phase transition. For instance, the density of states  as a function of real frequencies $\omega^\prime$, $N(\omega^\prime) \sim 1/[(\omega^\prime)^2 - \Delta(\omega^\prime)^2]$ will scale as $N(\omega^\prime) \sim 1/\sqrt{\omega^\prime - \Delta(0)}$ as $f \rightarrow f_c$, like in the BCS case.

\begin{figure}
\centering
\includegraphics[width=\columnwidth]{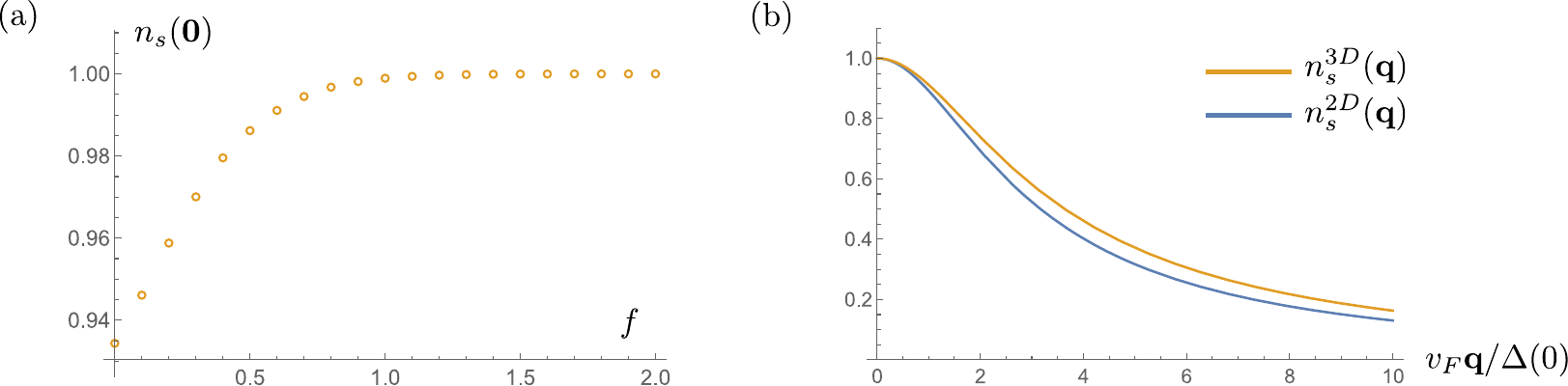}
\caption{Superfluid density. (a) Global superfluid density, at vanishing total momentum of a Cooper pair. (b) Momentum-dependent superfluid density, which parameterizes the phase stiffness at length scales $\sim 1/|\q|$. Fig.\ adapted from Ref.\ \cite{pimenov2021quantum}. }
\label{superfluidfig}
\end{figure}

 We note that the behavior $\Delta(0) \sim \exp(1/(f_c -f))$ can be interpreted as an infinite-order quantum phase transition. This is a result of the constant DoS at the Fermi level: in the low-density limit, $\Delta(0) \sim (f_c  - f)^2$ decays as a power law \cite{PhysRevB.105.064518}. The infinite order transition is not in the BKT universality class, which would correspond to $\exp(1/\sqrt{f - f_c})$ \cite{Kosterlitz_1974}. BKT-like transitions are associated with conformal invariance \cite{PhysRevD.80.125005}, while our interaction contains an explicit energy scale $\Omega_D$. On the other hand, the transition does share some common characteristics with the BKT transition. For instance, one can show that the superfluid density $n_s$, which measures the energy cost of spatial fluctuations,  has a universal jump at the transition \cite{pimenov2021quantum}, much like in the BKT case \cite{PhysRevLett.39.1201}: as the critical repulsion is approached from below, $f \nearrow f_c$, $n_s$ is locked to its BCS value (see Fig.\ \ref{superfluidfig} (a)).\footnote{Note that our system has both time-reversal and translational invariance, which are necessary conditions for $n_s =1$ (in proper units) according to Leggett's theorem \cite{Leggetttheorem}.} This means that the global phase of the BCS gap function is well-defined all the way up to the transition. On the other hand, one can define a momentum-dependent superfluid density $n_s(\q)$, which parametrizes the phase stiffness for length scales $r \sim 1/|\q|$. In both $2D$ and $3D$, this quantity is a scaling function of $v_F|\q|/\Delta(0)$, and it vanishes for $v_F |\q| \gg \Delta(0)$, see Fig. \ref{superfluidfig}(b).   This implies that, as the transition is approached, strong phase fluctuations occur on larger and larger length scales $\sim 1/\Delta(0)$ \cite{SchmalianComment}.

{Similar to the Rietschel-Scham model, the expressions for $\lambda_c, f_c$ can also be obtained within an effective low-energy model, which is valid for $\omega < \omega_0$  (cf.  Eq.\ \eqref{RSDelta}).
    Again, the coupling in such an effective model, $\lambda_\text{eff} \sim f - f_c$ changes sign at the transition, and the critical exponents are BCS-like. However, as $f \rightarrow f_c$, the window,
  where the low-energy description is valid, \textit{vanishes}
     because $\omega_0$ tends to zero. Therefore, the 
      effective low-energy description at $\omega < \omega_0$ cannot capture any finite energy properties such as excitations above the ground state.  
       In this respect, the HEF model differs from the Rietschel-Scham model, where the corresponding scale $\omega_1$ remains finite when $T_c$ vanishes. 
   }

 The peculiar interplay between $\omega_0$, $\Delta (0)$,  and $\Delta (\omega \geq \omega_0)$
 at $f \to f_c$ leads to a highly non-trivial solution of the
  linearized gap equation for infinitesimally small $\Delta (\omega)$  (point D in Fig.\ \ref{phasediag1}). The linearized gap equation at $T=0$ reads:
\begin{align}
\label{lineareq}
\Delta(\omega) = - \lambda_c  \int_{-\Lambda}^\Lambda d\omp \frac{\Delta(\omega')}
{|\omp|} \left( f - \frac{1}{1+(\omega - \omp)^2}\right)\ .
\end{align}
The right hand side is free from singularities if $\Delta(0) = 0$. This is expected
  as  $\omega_0 =0$ at $f = f_c$. Naively one should then search for a solution of the form  $\Delta(\omega) \sim \omega^2$
  for  $\omega \ll 1$ and flat $\Delta (\omega)$ at larger $\omega$.   However, such
   $\Delta(\omega)$ is sign-preserving, and, as just mentioned, there
   is no solution of (\ref{lineareq}) for a sign-preserving gap function.

\begin{figure}
\centering
\includegraphics[width=\columnwidth]{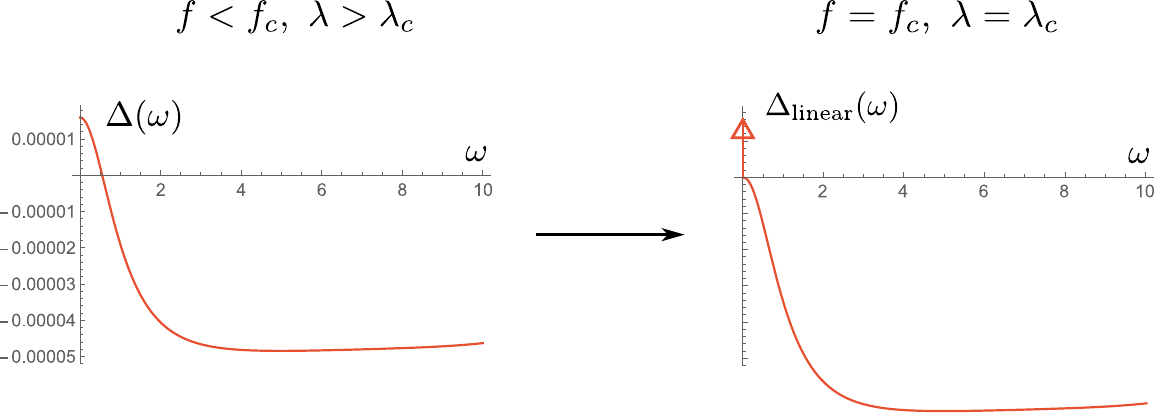}
\caption{Evolution of the solution $\Delta$ close to the transition line; note that the scale of the linear solution $\Delta_{\text{linear}}$ is not fixed.  The arrow denotes the delta-piece. }
\label{linfig}
\end{figure}

   A hint for the form of $\Delta (\omega)$ comes from the analysis of
    Eqs. (\ref{n1ref}) and (\ref{n2ref}) at small but finite $f_c -f$.  The gap function $\Delta (0)$ is parametrically smaller than $\Delta (\omega \geq \omega_0)$ and $\omega_0$, which determines the width of the range where the sign of $\Delta (\omega)$ is the same as of $\Delta (0)$, is also parametrically small. Yet, if we construct the integrals
     of $\Delta (\omega)/\sqrt{\omega^2 + \Delta^2 (\omega)}$ over the frequency range between $0$ and $\omega_0$ and over $\omega > \omega_0$,
         \begin{align}
          Y_1 &= \int_0^{\omega_0} \frac{\Delta (\omega)}{\sqrt{\omega^2 + \Delta^2 (\omega)}} \nonumber \\
     Y_2 &= \int_{\omega_0}^\Lambda \frac{\Delta (\omega)}{\sqrt{\omega^2 + \Delta^2 (\omega)}} ,
     \end{align}
We find that the ratio $Y_1/Y_2$ is independent of $f_c-f$.  This implies that as $f$ approaches $f_c$ from below and the magnitude of $\Delta (\omega)$ vanishes,
    $\Delta (\omega)/\sqrt{\omega^2 + \Delta^2 (\omega)} \approx \Delta (\omega)/\omega$ becomes $\gamma \delta (\omega)$,
     where $\gamma$ is of order $\Delta (\omega \geq \omega_0)$.  Then  the ratio $Y_2/Y_1$ remains finite at $f \to f_c$.
     We illustrate this in  Fig.\ \ref{linfig}, where we present the numerical solution of the non-linear gap equation at $f \to f_c$.

Using this as an input, we search for the solution of the linearized gap equation in the form
\begin{align}
\Delta(\omega) = \tilde\Delta(\omega) + \gamma \times |\omega|  \delta(\omega) ,
\label{aa}
\end{align}
where $\tilde\Delta(\omega)$ is a regular function, which scales as $\omega^2$ at small $\omega$.

It is convenient to introduce $D(\omega) = \Delta(\omega)/\omega$, which is an odd function of  $\omega$. In terms of $D(\omega)$, Eq.\ \eqref{lineareq} reads
\begin{align}
\label{Domega}
D(\omega) \omega = - \lambda_c \int_{-\Lambda}^\Lambda d\omp \sign(\omp) D(\omp) \left( f - \frac{1}{1+(\omega - \omp)^2} \right) \ .
\end{align}

 The ansatz for the gap function in terms of $D(\omega)$ is
\begin{align}
\label{Dansatz}
D(\omega) = \tilde D(\omega) + \gamma \delta(\omega) \sign(\omega) \ .
\end{align}
As a first step, we evaluate Eq.\ \eqref{Domega} at $\omega = 0+$, inserting \eqref{Dansatz} into \eqref{Domega}.  We obtain
\begin{align}
0 = - \lambda_c \int_{-\Lambda}^{\Lambda} d\omp \tilde D(\omp) \sign(\omp) \left( f - \frac{1}{1+(\omp)^2} \right) - \gamma \lambda_c \left ( f - \frac{1}{1+\omega^2}  \right) \ .
\end{align}
This equation again shows the necessity of the $\delta$-function term: it is required to cancel out the integral contribution from the sign-preserving regular part $\tilde D(\omp)$.

To find an analytical approximation to $\tilde D(\omega)$, we make an ansatz
\begin{align}
\label{tildeDansatz}
\tilde D(\omega) = \begin{cases} &- \frac{\omega}{\omega_o^2} \quad |\omega| < \omega_1 \\
&- \frac{1}{\omega} \quad |\omega| > \omega_1 \ ,
\end{cases}
\end{align}
where $\omega_1$ is a free parameter. Since the overall scale of the gap function is not fixed, we are free to  set  $\Delta(\Lambda) = 1$.

We insert the ansatz \eqref{tildeDansatz} into \eqref{Domega}, and derive three equations for $D(0)$, $D(\omega <
\omega_1)$ and $D(\omega
 > \omega_1)$. For large $\Lambda$, they can be   expressed as
\begin{align}
&\frac{1}{\lambda_c} = \gamma - \frac{\log(1 + \omega_1^2)}{\omega_1^2} - \log\left(1 + \frac{1}{\omega_1^2} \right)  \\ &
\frac{1}{\lambda_c} = \omega_1^2 \left[ \gamma + I_1 (\omega_1) + I_2(\omega_1) \right]  \\
&\frac{1}{\lambda_c} = f\left[\gamma - 1  - 2 \log \left( \frac{\Lambda}{\omega_1} \right) \right] , \\
& I_1(\omega_1) = \int_0^{\omega_1} dx \frac{x}{\omega_1^2} \left ( \frac{3x^2 -1}{(1+x^2)^3} \right),   \quad \quad  I_2(\omega_1) = \int_{\omega_1}^\Lambda dx \frac{1}{x} \left( \frac{3x^2 -1}{(1 + x^2)^3} \right) \ .
\end{align}
 At weak coupling ($\lambda_c  \ll 1$)
  these equations can easily be solved to leading order in $\lambda$ and yield
\begin{align}
\label{gammanalyt}
\gamma = \frac{1}{\lambda_c} , \quad \omega_1 = 1 , \quad  f = \frac{1}{1- 2\lambda_c \log\left(\Lambda \right) }
\end{align}
The value of $f$ agrees with our previous estimate \eqref{fcvalue}.

\begin{figure}
\centering
\includegraphics[width=.5\columnwidth]{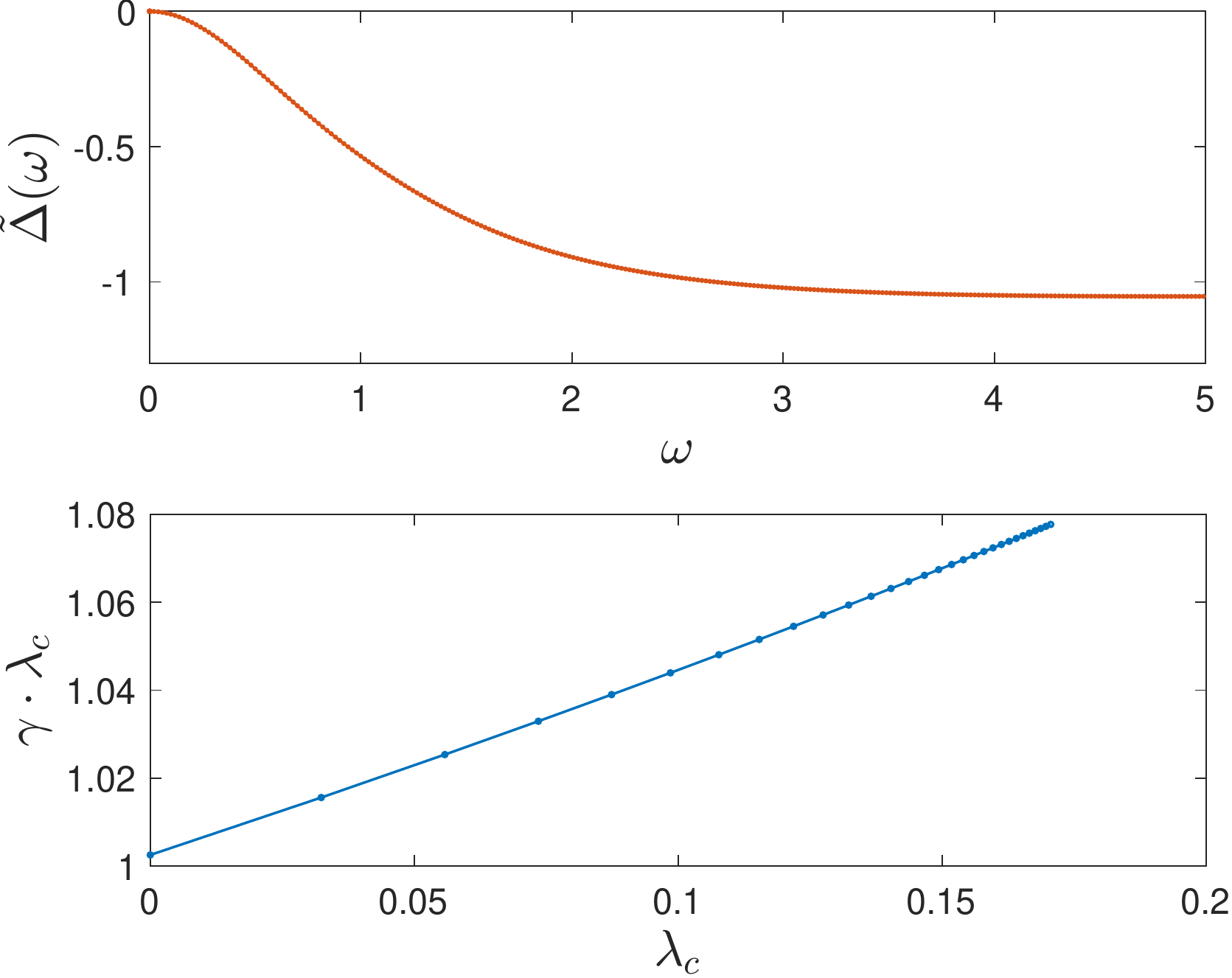}
\caption{Numerical solution of the linearized gap equation. Upper panel: regular part of the gap function $\tilde{\Delta}(\omega)$ for generic parameters. Lower panel: weight of the delta-function part as function of critical $\lambda_c$; }
\label{gammaplot}
\end{figure}

We can also determine $\gamma, \tilde \Delta$ by solving  Eq.\ \eqref{lineareq} numerically. We present the numerical solution in Fig.\ \ref{gammaplot}. In the upper panel of that figure we show the regular part $\tilde \Delta(\omega)$ for generic parameters. In the lower part we  show   $\gamma$ as a function of the critical $\lambda_c$. One can see that $\gamma \sim 1/\lambda_c$, as expected from Eq.\ \eqref{gammanalyt}.

\begin{figure}
\centering
\includegraphics[width=\columnwidth]{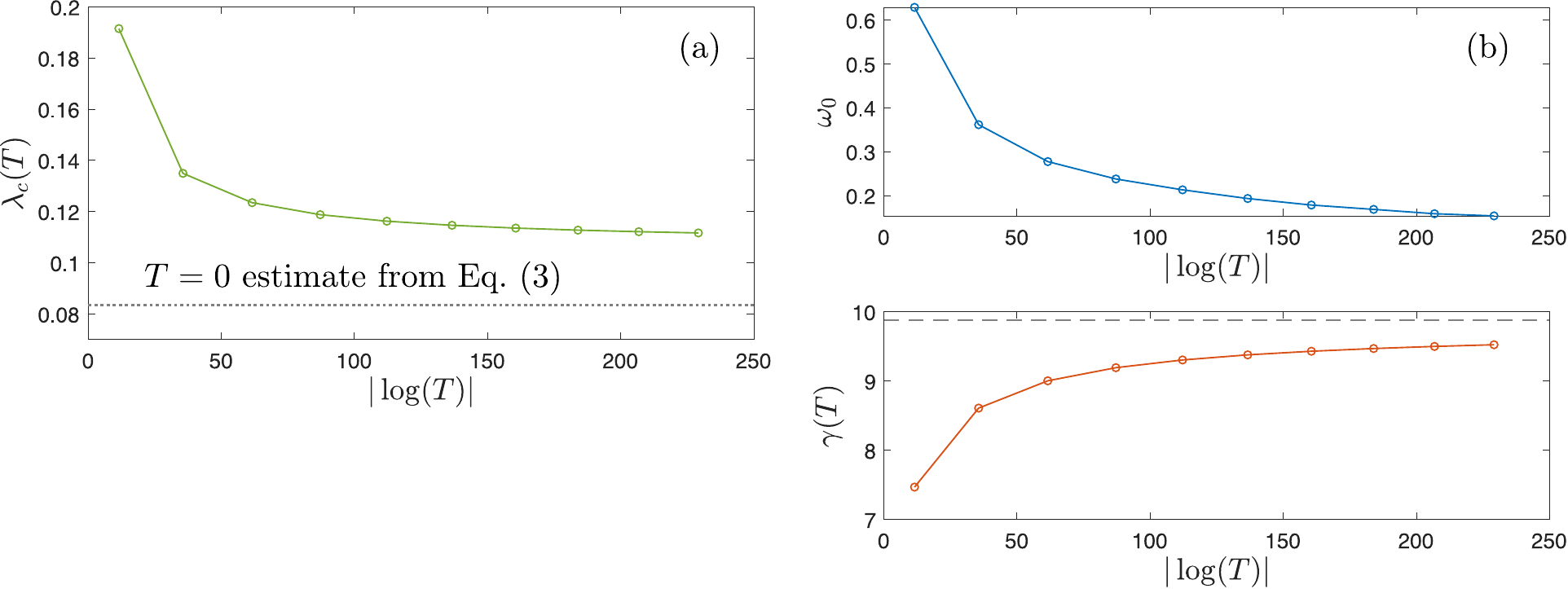}
\caption{Numerical solution of the linearized gap equation at a finite temperature, for parameters $\Lambda = 20, f = 2$. (a) Critical coupling for at finite temperature. The dotted line corresponds to the $T = 0$ estimate of Eq.\ \eqref{lambdac}, which holds with logarithmic accuracy. (b) Upper panel: position of the node.  Lower panel: integral weight $\gamma(T) \equiv 2\pi T \sum_{|\omega_m| < \omega_0} \Delta(\omega_m)/|\omega_m|$. As $T \rightarrow 0$, $\gamma(T)$ slowly approaches the zero-temperature weight (dashed line), which is taken from Fig.\ \ref{gammaplot}.}
\label{weight}
\end{figure}

Yet another way to check the appearance of the delta-function is by solving the linearized gap equation at a finite temperature. The result is shown in Fig.\ \ref{weight}. As $T \rightarrow 0$, the finite-temperature gap correctly approaches the zero-temperature result.

To the best of our knowledge, the appearance of the singular $\delta$-functional piece in the solution of the linearized gap equation has never before been discussed in the literature.  A further analysis of the critical exponents of this non-trivial phase transition, possibly using RG  techniques, is
    an interesting problem for future research.  In a two-dimensional system, the transition could also be probed experimentally by modifying the effective strength of the Coulomb repulsion via screening. This can be achieved via gating or changing the dielectric constant of a substrate \cite{SchmalianComment}.

{ We now argue that the presence of the $\delta$ function in $\Delta (\omega)$ at $f=f_c -0$ is the necessary consequence of the fact that the gap function at $f < f_c$ has a vortex on the Matsubara axis.  Indeed, expressing
 $\delta (\omega)$ as $(1/\pi) x/(x^2 +\omega^2)$, where $x$ is infinitesimally small,  and extending Eq. (\ref{aa}) into the complex plane by replacing $\omega$ by $-i z$, where $z = \omega' +i \omega^{''}$, we find that $\Delta (z)$ has
  zeros (vortex points) along the Matsubara axis,  at $z = \pm i b x^{1/3}$, $b = O(1)$, and poles (antivortex points) infinitesimally below the real axis, at $z = \pm x -i0$.  This clearly implies that the appearance of a superconducting state at $f < f_c$ with sign-changing gap along the Matsubara axis is the result of an unbinding of two vortex-antivortex pairs. The pairs are located at $z =0$ at $f = f_c$. After unbinding, vortices move along the Matsubara axis and anti-vortices move into the lower half-plane of complex frequency.   We consider this as a strong evidence that the transition at $f =f_c$ is  at least partly  described by BKT-like physics. }

\section{Odd-frequency gap function}
\label{OFsec}

Another potential option for superconductivity from a repulsive interaction is an OF gap function
 $\Delta (\omega) = - \Delta (-\omega)$.  Such pairing is not forbidden on general grounds, as was first
  recognized  by Berezinskii \cite{berezinskii1974new}, but to satisfy  the Pauli principle, an OF gap function must either be odd in momentum space or be in the spin-triplet channel. In our case the interaction $V(\Omega)$ does not depend on momentum, hence odd-frequency gap function must be a spin triplet.

 OF superconductivity is a rich field which has been theoretically studied for decades. We refer a reader to Refs.\ \cite{Kivelson_1992,Golubov_2009, doi:10.1143/JPSJ.81.011013, RevModPhys.91.045005} for theoretical reasoning
  and to Refs.\ \cite{DiBernardo2015a, PhysRevX.5.041021} for the discussion of a potential experimental observation of OF superconductivity in heterostructures.

In our case, at small $f <1$,  OF pairing is prevented by the development of EF pairing at much a higher $T$ because OF pairing is a threshold phenomenon (see below), while EF pairing is not. However, at larger $f$ the situation may change as the Hubbard repulsion, which acts against EF pairing,  cancels out in the odd-frequency channel~\cite{PhysRevB.45.13125, doi:10.1143/JPSJ.80.044711, PhysRevB.100.180502}.  We show below that the actual situation is more involved.

\begin{figure}
\centering
\includegraphics[width=\columnwidth]{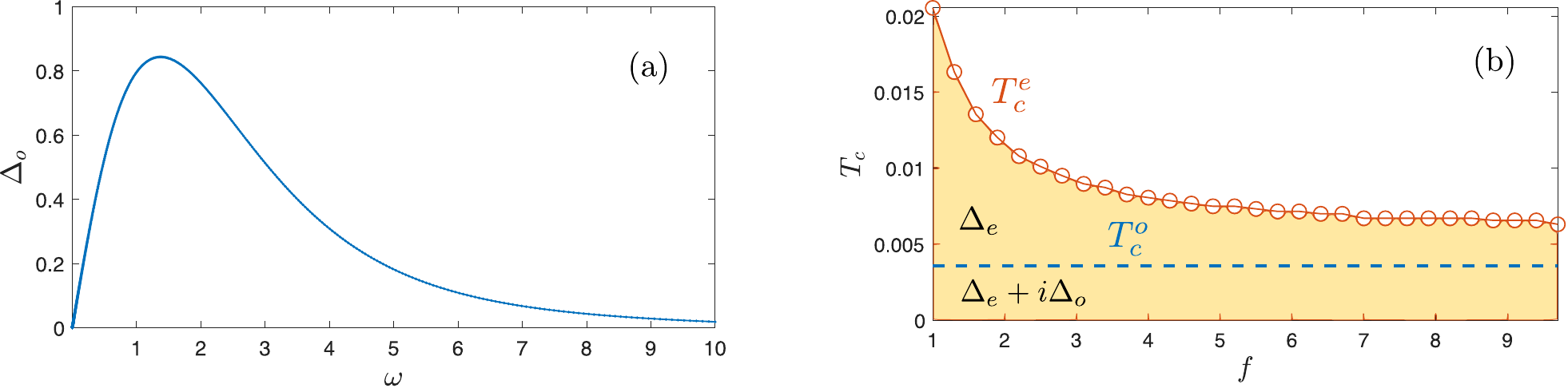}
\caption{(a) OF gap function, denoted as $\Delta_o$, at $T = 0$ for generic parameters. To obtain it, we  neglected the self-energy. (b) Exemplary phase diagram, taken from \cite{pimenov2022odd}. $\Delta_e$ denotes the EF solution, $\Delta_e + i \Delta_o$ the mixed state with breaking of time-reversal symmetry.}
\label{OFplot}
\end{figure}

Indeed, suppose momentarily that EF superconductivity does not develop.  The  gap equation for the OF $\Delta_o (\omega)$ is
\begin{align}
&\label{ofgapeq}  \Delta_o(\omega) = - \frac{1}{2}
 \int_{-\Lambda}^\Lambda d\omp \frac{\chi_o(\omega, \omp)}{\sqrt{(\omp)^2 + |\Delta_o (\omp)|^2}  } \times \Delta_o(\omp)   \\  &V_o(\omega, \omp) = \chi(\omega - \omp) - \chi(\omega + \omp) = - \frac{4\lambda\times \omega \omp}{\left( 1+ (\omega - \omp)^2 \right) \left( 1+ (\omega + \omp)^2 \right) }  \ .
\end{align}
We see that  $\chi_o (\omega, \omp)$ scales linearly with $\omp$. The r.h.s.\ of (\ref{ofgapeq}) then does not contain a Cooper logarithm
\cite{RevModPhys.91.045005},
 hence a non-zero $\Delta_o (\omega)$ appears only when $\lambda$ exceeds a certain threshold $\lambda_o$. To judge whether EF  or OF pairing
 develops  for a given set of parameters, one must compare $\lambda_o$ with the threshold coupling for EF pairing
 $\lambda_e = \lambda_c  \sim (f-1)/(2f\log(\Lambda))$ from Eq.\ \eqref{lambdac}.
   By standard reasoning, the order for which the threshold value of $\lambda$ is smaller, develops.
  This analysis suggests that favorable parameters for OF pairing are (i)
  a large repulsion $f$ and (ii) a fairly small UV cutoff ($\sim$ Fermi energy) $\Lambda$. An exemplary solution for an odd-frequency gap function is shown in Fig.\ \ref{OFplot}(a).

However, there is another caveat. In the analysis above we neglected the fermionic self-energy. This is justified at weak coupling, but not for  $\lambda > \lambda_o$.  A self-energy makes fermions less coherent and therefore acts against pairing~\cite{PhysRevB.47.513, langmannOF}. Within Eliashberg theory, the same interaction that contributes to pairing also gives rise to the self-energy. The theory then yields a set of two coupled equations for the pairing vertex
  $\Phi$ and the self-energy $\Sigma$.  The gap function $\Delta (\omega)$ is expressed via $\Phi$ and $\Sigma$  as
 $\Delta(\omega) = \Phi(\omega)/(1 + \Sigma(\omega)/\omega)$  (for vanishing $\Sigma$, $\Delta = \Phi$).
  The two Eliashberg equations for $\Phi$ and $\Sigma$ can be re-expressed as equations for
 $\Delta$ and the inverse fermionic residue $Z (\omega) = 1 + \Sigma(\omega)/\omega$, and the equation for
 $\Delta (\omega)$ contains only $\Delta (\omega')$. For the OF gap function the full gap equation is
\begin{align}
\label{ofgapeqwithSigma}
\Delta_o(\omega) = -\frac{1}{2} \int_{-\Lambda}^\Lambda d\omp \frac{\chi_o(\omega, \omp)}{\sqrt{(\omp)^2 + |\Delta_o(\omp)|^2}  } \times \left ( \Delta_o(\omp)  -  \frac{\omp}{\omega}  \Delta_o(\omega) \right) \ .
 \end{align}
The self-energy contributes the part $\sim \frac{\omp}{\omega} \Delta(\omega)$ on the r.h.s.\ of Eq.\ \eqref{ofgapeqwithSigma}. Since $\Delta_o(0) = 0$, one can linearize Eq.~\eqref{ofgapeqwithSigma} without encountering singularities.  Solving the latter one can verify how the self-energy affects the critical $\lambda_o$.
 The result is somewhat unexpected: the critical $\lambda_o$ becomes infinite, i.e., for any finite $\lambda$ OF superconductivity does not develop \cite{pimenov2022odd, langmannOF}.

For the EF gap function, the self-energy correction is still proportional to $V_o$, and explicit calculation show that it only slightly shifts the threshold value $\lambda_e$.

At this point, the prospects of achieving OF superconductivity seem to be dire. However, we recall that an infinite threshold $\lambda_o$ is obtained assuming that the interaction in the particle-hole channel is the same one as in the particle-particle channel. This holds in the Eliashberg approximation, but does not hold beyond it, when one includes vertex corrections~\cite{PhysRevB.104.174518, pimenov2022odd}. It turns out that vertex corrections  suppress
the interaction in the particle-hole channel relative to the one in the particle-particle channel.
In one-loop order for vertex corrections  the suppression factor is
\begin{align}
\alpha \sim  \frac{1 + 2\lambda f}{1 + 4\lambda f} < 1 \ .
\end{align}
Note that this vertex correction does not carry the Migdal parameter $\Omega_D/E_F$. This has been discussed in
 e.g., Ref. \cite{ALLEN19831}.

 We remark that the correct evaluation of vertex corrections is somewhat tricky at finite temperatures, as even in the presence of vertex corrections the self-energy still exactly cancels out the thermal contribution from $\omega' =\omega$ on the r.h.s.\ of the gap equation (see Ref.\   \cite{pimenov2022odd} for details).

 For $\alpha<1$, $\lambda_o$  becomes finite, and when $\alpha$ is small enough,
  $\lambda_o$ and $\lambda_e$ become comparable.  Explicit calculations show~\cite{pimenov2022odd} that
$\lambda_e$ is smaller for all $\alpha \geq 0$, unless extreme parameters are chosen. Hence, the system first develops EF pairing at a certain temperature $T_c^e$. However, because $\lambda_e$ and $\lambda_o$  are comparable,
the OF component does develop at a lower temperature $T_c^o$, and below this $T$ the order parameter contains both even and odd components.

 Interestingly, the relative phase between the two is $\pm \pi/2$
 \cite{PhysRevB.104.174518, pimenov2022odd}, i.e.,
\begin{align}
\label{mixed}
\Delta(\omega) = \Delta_e + i \Delta_o(\omega) \quad \text{or} \quad \Delta(\omega) = \Delta_e - i \Delta_o(\omega)
\end{align}
The system spontaneously chooses one of the two orders, and in this way \textit{spontaneously} breaks time-reversal invariance (which acts on gap functions on the Matsubara axis simply as complex conjugation). This is a rare example of time-reversal symmetry breaking in a one-band $s$-wave superconductor. Detection of time-reversal symmetry breaking can be achieved with methods such as Kerr rotation or muon spin relaxation \cite{Ghosh_2020}. We show a phase diagram including the  mixed-state region in Fig.\ \ref{OFplot}(b).

Since the gap function in \eqref{mixed} is complex, $\Delta (\omega) = |\Delta (\omega)|e^{i\eta (\omega)}$,
 it does not have nodes. Nodal points, however, may exist in the upper half-plane of a complex frequency, each of them a center of a dynamical vortex.  Note also that $\eta (\omega)$ is odd in $\omega$, hence
  $\Delta (\omega' + i\omega)$ and $\Delta (\omega' -i\omega)$ are not equivalent, unlike the case shown in Fig.\ \ref{Morten1}.

\section{Summary and open questions}
\label{concsec}
In this paper we  reviewed some recent progress in the study of  an $s-$wave
  superconductivity from a dynamical  repulsive interaction.

  We argued that superconductivity is possible even if the pairing interaction is repulsive at all frequencies,
  but the gap $\Delta (\omega)$ necessary changes sign at least once along a positive Matsubara axis.
   We related nodal points of $\Delta (\omega)$ to dynamical vortices and argued that their presence gives rise to
     extra variation of the phase of a complex gap function between large negative and large positive real frequencies, in multiples of $2\pi$. We argued that $s-$wave superconductivity out of repulsion is a threshold phenomenon, and studied in detail the quantum phase transition at $T=0$ in which a superconducting order parameter vanishes.
This transition is highly non-trivial as both the gap amplitude and the frequency, at which the gap changes sign, vanish at criticality in a particular manner.  As a consequence, the gap function at an infinitesimal distance from the transition has a singular $\delta-$functional piece. Without this piece, one cannot obtain a solution of the linearized gap equation at the critical point. Finally, we argued that for a repulsive interaction there is a competition between EF and OF pairing
  as the latter is not affected by Hubbard repulsion. We showed that the likely outcome of this competition is a mixed state in which EF and OF gap components are both nonzero. The relative phase between the two is $\pm \pi/2$. The system spontaneously chooses $\pi/2$ or $-\pi/2$ and in this way spontaneously breaks time-reversal symmetry.

 There are several  open questions and ongoing challenges in the field. First, it would be advantageous to identify physical observables which could detect dynamical vortices without the need to sample the full real frequency axis.
 One possible place to look is the transient response of a superconductor, for instance if the repulsion $f$ is changed in time. Second, it would be interesting to extend the present approach to low-density and flat band systems \cite{PhysRevB.105.064518, PhysRevB.102.235423, doi:10.1126/sciadv.abh2233, PhysRevB.105.094506}. Third, an issue to consider is whether one can construct a quantum simulator of the electron-phonon systems, which offers more experimental flexibility. Cavity systems, where retarded interaction can be mediated by massive cavity photons, could be a viable candidate \cite{PhysRevLett.122.133602}. Fourth, it would be interesting to obtain critical exponents of the superconductor to normal-state transition.   Fifth, several groups recently discussed sign-changing gap functions $\Delta (\omega)$ for  superconductivity at a quantum-critical point towards some particle-hole order~~\cite{abanov}
       and for superconductivity in SYK-type models~\cite{laura}. It would be interesting to compare these gap functions with the ones we studied here.
       Sixth,
         a challenging question is whether the pairing fluctuation propagator, which becomes gapless at the HEF model quantum phase transition, leads to non-Fermi-liquid scattering rate of fermions.
Finally, it would be interesting to address the issue whether one can construct analogous theories for other instabilities besides superconductivity.

\section*{Acknowledgement}

We thank
A. Balatsky, S. Kivelson, E. Langmann,  J. Schmalian, S.-S.\ Zhang for useful discussions and
  suggestions.
The work by A.V.C  was supported by the NSF DMR-1834856.
A.V.C acknowledges the hospitality of KITP at UCSB, where part of the
work has been conducted. The research at KITP is supported by the
National Science Foundation under Grant No. NSF PHY-1748958.

\bibliography{repReview}

\begin{thebibliography}{59}%
\makeatletter
\providecommand \@ifxundefined [1]{%
 \@ifx{#1\undefined}
}%
\providecommand \@ifnum [1]{%
 \ifnum #1\expandafter \@firstoftwo
 \else \expandafter \@secondoftwo
 \fi
}%
\providecommand \@ifx [1]{%
 \ifx #1\expandafter \@firstoftwo
 \else \expandafter \@secondoftwo
 \fi
}%
\providecommand \natexlab [1]{#1}%
\providecommand \enquote  [1]{``#1''}%
\providecommand \bibnamefont  [1]{#1}%
\providecommand \bibfnamefont [1]{#1}%
\providecommand \citenamefont [1]{#1}%
\providecommand \href@noop [0]{\@secondoftwo}%
\providecommand \href [0]{\begingroup \@sanitize@url \@href}%
\providecommand \@href[1]{\@@startlink{#1}\@@href}%
\providecommand \@@href[1]{\endgroup#1\@@endlink}%
\providecommand \@sanitize@url [0]{\catcode `\\12\catcode `\$12\catcode
  `\&12\catcode `\#12\catcode `\^12\catcode `\_12\catcode `\%12\relax}%
\providecommand \@@startlink[1]{}%
\providecommand \@@endlink[0]{}%
\providecommand \url  [0]{\begingroup\@sanitize@url \@url }%
\providecommand \@url [1]{\endgroup\@href {#1}{\urlprefix }}%
\providecommand \urlprefix  [0]{URL }%
\providecommand \Eprint [0]{\href }%
\providecommand \doibase [0]{https://doi.org/}%
\providecommand \selectlanguage [0]{\@gobble}%
\providecommand \bibinfo  [0]{\@secondoftwo}%
\providecommand \bibfield  [0]{\@secondoftwo}%
\providecommand \translation [1]{[#1]}%
\providecommand \BibitemOpen [0]{}%
\providecommand \bibitemStop [0]{}%
\providecommand \bibitemNoStop [0]{.\EOS\space}%
\providecommand \EOS [0]{\spacefactor3000\relax}%
\providecommand \BibitemShut  [1]{\csname bibitem#1\endcsname}%
\let\auto@bib@innerbib\@empty
\bibitem [{\citenamefont {Bardeen}\ \emph {et~al.}(1957)\citenamefont
  {Bardeen}, \citenamefont {Cooper},\ and\ \citenamefont
  {Schrieffer}}]{PhysRev.106.162}%
  \BibitemOpen
  \bibfield  {author} {\bibinfo {author} {\bibfnamefont {J.}~\bibnamefont
  {Bardeen}}, \bibinfo {author} {\bibfnamefont {L.~N.}\ \bibnamefont
  {Cooper}},\ and\ \bibinfo {author} {\bibfnamefont {J.~R.}\ \bibnamefont
  {Schrieffer}},\ }\href {https://doi.org/10.1103/PhysRev.106.162} {\bibfield
  {journal} {\bibinfo  {journal} {Phys. Rev.}\ }\textbf {\bibinfo {volume}
  {106}},\ \bibinfo {pages} {162} (\bibinfo {year} {1957})}\BibitemShut
  {NoStop}%
\bibitem [{\citenamefont {Cooper}(1956)}]{PhysRev.104.1189}%
  \BibitemOpen
  \bibfield  {author} {\bibinfo {author} {\bibfnamefont {L.~N.}\ \bibnamefont
  {Cooper}},\ }\href {https://doi.org/10.1103/PhysRev.104.1189} {\bibfield
  {journal} {\bibinfo  {journal} {Phys. Rev.}\ }\textbf {\bibinfo {volume}
  {104}},\ \bibinfo {pages} {1189} (\bibinfo {year} {1956})}\BibitemShut
  {NoStop}%
\bibitem [{\citenamefont {Kohn}\ and\ \citenamefont
  {Luttinger}(1965)}]{PhysRevLett.15.524}%
  \BibitemOpen
  \bibfield  {author} {\bibinfo {author} {\bibfnamefont {W.}~\bibnamefont
  {Kohn}}\ and\ \bibinfo {author} {\bibfnamefont {J.~M.}\ \bibnamefont
  {Luttinger}},\ }\href {https://doi.org/10.1103/PhysRevLett.15.524} {\bibfield
   {journal} {\bibinfo  {journal} {Phys. Rev. Lett.}\ }\textbf {\bibinfo
  {volume} {15}},\ \bibinfo {pages} {524} (\bibinfo {year} {1965})}\BibitemShut
  {NoStop}%
\bibitem [{\citenamefont {Shankar}(1994)}]{RevModPhys.66.129}%
  \BibitemOpen
  \bibfield  {author} {\bibinfo {author} {\bibfnamefont {R.}~\bibnamefont
  {Shankar}},\ }\href {https://doi.org/10.1103/RevModPhys.66.129} {\bibfield
  {journal} {\bibinfo  {journal} {Rev. Mod. Phys.}\ }\textbf {\bibinfo {volume}
  {66}},\ \bibinfo {pages} {129} (\bibinfo {year} {1994})}\BibitemShut
  {NoStop}%
\bibitem [{\citenamefont {Maiti}\ and\ \citenamefont
  {Chubukov}(2013)}]{doi:10.1063/1.4818400}%
  \BibitemOpen
  \bibfield  {author} {\bibinfo {author} {\bibfnamefont {S.}~\bibnamefont
  {Maiti}}\ and\ \bibinfo {author} {\bibfnamefont {A.~V.}\ \bibnamefont
  {Chubukov}},\ }\href@noop {} {\bibfield  {journal} {\bibinfo  {journal} {AIP
  Conference Proceedings}\ }\textbf {\bibinfo {volume} {1550}},\ \bibinfo
  {pages} {3} (\bibinfo {year} {2013})}\BibitemShut {NoStop}%
\bibitem [{\citenamefont {Kagan}\ \emph {et~al.}(2014)\citenamefont {Kagan},
  \citenamefont {Val'kov}, \citenamefont {Mitskan},\ and\ \citenamefont
  {Korovushkin}}]{KaganKL}%
  \BibitemOpen
  \bibfield  {author} {\bibinfo {author} {\bibfnamefont {M.~Y.}\ \bibnamefont
  {Kagan}}, \bibinfo {author} {\bibfnamefont {V.~V.}\ \bibnamefont {Val'kov}},
  \bibinfo {author} {\bibfnamefont {V.~A.}\ \bibnamefont {Mitskan}},\ and\
  \bibinfo {author} {\bibfnamefont {M.~M.}\ \bibnamefont {Korovushkin}},\
  }\href {https://doi.org/10.1134/S1063776114060132} {\bibfield  {journal}
  {\bibinfo  {journal} {Journal of Experimental and Theoretical Physics}\
  }\textbf {\bibinfo {volume} {118}},\ \bibinfo {pages} {995} (\bibinfo {year}
  {2014})}\BibitemShut {NoStop}%
\bibitem [{\citenamefont {Tolmachev}\ and\ \citenamefont
  {Tiablikov}(1958)}]{tolmachev1958new}%
  \BibitemOpen
  \bibfield  {author} {\bibinfo {author} {\bibfnamefont {V.~V.}\ \bibnamefont
  {Tolmachev}}\ and\ \bibinfo {author} {\bibfnamefont {S.~V.}\ \bibnamefont
  {Tiablikov}},\ }\href@noop {} {\bibfield  {journal} {\bibinfo  {journal}
  {Soviet Physics JETP}\ }\textbf {\bibinfo {volume} {34}} (\bibinfo {year}
  {1958})}\BibitemShut {NoStop}%
\bibitem [{\citenamefont {Bogoljubov}\ \emph {et~al.}(1958)\citenamefont
  {Bogoljubov}, \citenamefont {Tolmachov},\ and\ \citenamefont
  {Širkov}}]{Bogoljubov1958}%
  \BibitemOpen
  \bibfield  {author} {\bibinfo {author} {\bibfnamefont {N.~N.}\ \bibnamefont
  {Bogoljubov}}, \bibinfo {author} {\bibfnamefont {V.~V.}\ \bibnamefont
  {Tolmachov}},\ and\ \bibinfo {author} {\bibfnamefont {D.~V.}\ \bibnamefont
  {Širkov}},\ }\href
  {https://doi.org/https://doi.org/10.1002/prop.19580061102} {\bibfield
  {journal} {\bibinfo  {journal} {Fortschritte der Physik}\ }\textbf {\bibinfo
  {volume} {6}},\ \bibinfo {pages} {605} (\bibinfo {year} {1958})}\BibitemShut
  {NoStop}%
\bibitem [{\citenamefont {McMillan}(1968)}]{PhysRev.167.331}%
  \BibitemOpen
  \bibfield  {author} {\bibinfo {author} {\bibfnamefont {W.~L.}\ \bibnamefont
  {McMillan}},\ }\href {https://doi.org/10.1103/PhysRev.167.331} {\bibfield
  {journal} {\bibinfo  {journal} {Phys. Rev.}\ }\textbf {\bibinfo {volume}
  {167}},\ \bibinfo {pages} {331} (\bibinfo {year} {1968})}\BibitemShut
  {NoStop}%
\bibitem [{\citenamefont {Scalapino}\ \emph {et~al.}(1966)\citenamefont
  {Scalapino}, \citenamefont {Schrieffer},\ and\ \citenamefont
  {Wilkins}}]{PhysRev.148.263}%
  \BibitemOpen
  \bibfield  {author} {\bibinfo {author} {\bibfnamefont {D.~J.}\ \bibnamefont
  {Scalapino}}, \bibinfo {author} {\bibfnamefont {J.~R.}\ \bibnamefont
  {Schrieffer}},\ and\ \bibinfo {author} {\bibfnamefont {J.~W.}\ \bibnamefont
  {Wilkins}},\ }\href {https://doi.org/10.1103/PhysRev.148.263} {\bibfield
  {journal} {\bibinfo  {journal} {Phys. Rev.}\ }\textbf {\bibinfo {volume}
  {148}},\ \bibinfo {pages} {263} (\bibinfo {year} {1966})}\BibitemShut
  {NoStop}%
\bibitem [{\citenamefont {Morel}\ and\ \citenamefont
  {Anderson}(1962)}]{PhysRev.125.1263}%
  \BibitemOpen
  \bibfield  {author} {\bibinfo {author} {\bibfnamefont {P.}~\bibnamefont
  {Morel}}\ and\ \bibinfo {author} {\bibfnamefont {P.~W.}\ \bibnamefont
  {Anderson}},\ }\href {https://doi.org/10.1103/PhysRev.125.1263} {\bibfield
  {journal} {\bibinfo  {journal} {Phys. Rev.}\ }\textbf {\bibinfo {volume}
  {125}},\ \bibinfo {pages} {1263} (\bibinfo {year} {1962})}\BibitemShut
  {NoStop}%
\bibitem [{\citenamefont {Carbotte}(1990)}]{RevModPhys.62.1027}%
  \BibitemOpen
  \bibfield  {author} {\bibinfo {author} {\bibfnamefont {J.~P.}\ \bibnamefont
  {Carbotte}},\ }\href {https://doi.org/10.1103/RevModPhys.62.1027} {\bibfield
  {journal} {\bibinfo  {journal} {Rev. Mod. Phys.}\ }\textbf {\bibinfo {volume}
  {62}},\ \bibinfo {pages} {1027} (\bibinfo {year} {1990})}\BibitemShut
  {NoStop}%
\bibitem [{\citenamefont {Gurevich}\ \emph {et~al.}(1962)\citenamefont
  {Gurevich}, \citenamefont {Larkin},\ and\ \citenamefont
  {Firsov}}]{gurevich1962possibility}%
  \BibitemOpen
  \bibfield  {author} {\bibinfo {author} {\bibfnamefont {V.}~\bibnamefont
  {Gurevich}}, \bibinfo {author} {\bibfnamefont {A.}~\bibnamefont {Larkin}},\
  and\ \bibinfo {author} {\bibfnamefont {Y.~A.}\ \bibnamefont {Firsov}},\
  }\href@noop {} {\bibfield  {journal} {\bibinfo  {journal} {Sov. Phys.-Solid
  State (Engl. Transl.);(United States)}\ }\textbf {\bibinfo {volume} {4}}
  (\bibinfo {year} {1962})}\BibitemShut {NoStop}%
\bibitem [{\citenamefont {Rietschel}\ and\ \citenamefont
  {Sham}(1983)}]{PhysRevB.28.5100}%
  \BibitemOpen
  \bibfield  {author} {\bibinfo {author} {\bibfnamefont {H.}~\bibnamefont
  {Rietschel}}\ and\ \bibinfo {author} {\bibfnamefont {L.~J.}\ \bibnamefont
  {Sham}},\ }\href {https://doi.org/10.1103/PhysRevB.28.5100} {\bibfield
  {journal} {\bibinfo  {journal} {Phys. Rev. B}\ }\textbf {\bibinfo {volume}
  {28}},\ \bibinfo {pages} {5100} (\bibinfo {year} {1983})}\BibitemShut
  {NoStop}%
\bibitem [{\citenamefont {Ruhman}\ and\ \citenamefont
  {Lee}(2016)}]{PhysRevB.94.224515}%
  \BibitemOpen
  \bibfield  {author} {\bibinfo {author} {\bibfnamefont {J.}~\bibnamefont
  {Ruhman}}\ and\ \bibinfo {author} {\bibfnamefont {P.~A.}\ \bibnamefont
  {Lee}},\ }\href {https://doi.org/10.1103/PhysRevB.94.224515} {\bibfield
  {journal} {\bibinfo  {journal} {Phys. Rev. B}\ }\textbf {\bibinfo {volume}
  {94}},\ \bibinfo {pages} {224515} (\bibinfo {year} {2016})}\BibitemShut
  {NoStop}%
\bibitem [{\citenamefont {Ruhman}\ and\ \citenamefont
  {Lee}(2017)}]{PhysRevB.96.235107}%
  \BibitemOpen
  \bibfield  {author} {\bibinfo {author} {\bibfnamefont {J.}~\bibnamefont
  {Ruhman}}\ and\ \bibinfo {author} {\bibfnamefont {P.~A.}\ \bibnamefont
  {Lee}},\ }\href {https://doi.org/10.1103/PhysRevB.96.235107} {\bibfield
  {journal} {\bibinfo  {journal} {Phys. Rev. B}\ }\textbf {\bibinfo {volume}
  {96}},\ \bibinfo {pages} {235107} (\bibinfo {year} {2017})}\BibitemShut
  {NoStop}%
\bibitem [{\citenamefont {Chubukov}\ \emph {et~al.}(2019)\citenamefont
  {Chubukov}, \citenamefont {Prokof'ev},\ and\ \citenamefont
  {Svistunov}}]{PhysRevB.100.064513}%
  \BibitemOpen
  \bibfield  {author} {\bibinfo {author} {\bibfnamefont {A.}~\bibnamefont
  {Chubukov}}, \bibinfo {author} {\bibfnamefont {N.~V.}\ \bibnamefont
  {Prokof'ev}},\ and\ \bibinfo {author} {\bibfnamefont {B.~V.}\ \bibnamefont
  {Svistunov}},\ }\href {https://doi.org/10.1103/PhysRevB.100.064513}
  {\bibfield  {journal} {\bibinfo  {journal} {Phys. Rev. B}\ }\textbf {\bibinfo
  {volume} {100}},\ \bibinfo {pages} {064513} (\bibinfo {year}
  {2019})}\BibitemShut {NoStop}%
\bibitem [{\citenamefont {W\"olfle}\ and\ \citenamefont
  {Balatsky}(2018)}]{PhysRevB.98.104505}%
  \BibitemOpen
  \bibfield  {author} {\bibinfo {author} {\bibfnamefont {P.}~\bibnamefont
  {W\"olfle}}\ and\ \bibinfo {author} {\bibfnamefont {A.~V.}\ \bibnamefont
  {Balatsky}},\ }\href {https://doi.org/10.1103/PhysRevB.98.104505} {\bibfield
  {journal} {\bibinfo  {journal} {Phys. Rev. B}\ }\textbf {\bibinfo {volume}
  {98}},\ \bibinfo {pages} {104505} (\bibinfo {year} {2018})}\BibitemShut
  {NoStop}%
\bibitem [{\citenamefont {Eliashberg}(1960)}]{eliashberg1960interactions}%
  \BibitemOpen
  \bibfield  {author} {\bibinfo {author} {\bibfnamefont {G.}~\bibnamefont
  {Eliashberg}},\ }\href@noop {} {\bibfield  {journal} {\bibinfo  {journal}
  {Sov. Phys. JETP}\ }\textbf {\bibinfo {volume} {11}},\ \bibinfo {pages} {696}
  (\bibinfo {year} {1960})}\BibitemShut {NoStop}%
\bibitem [{\citenamefont {Marsiglio}(2020)}]{MARSIGLIO2020168102}%
  \BibitemOpen
  \bibfield  {author} {\bibinfo {author} {\bibfnamefont {F.}~\bibnamefont
  {Marsiglio}},\ }\href
  {https://doi.org/https://doi.org/10.1016/j.aop.2020.168102} {\bibfield
  {journal} {\bibinfo  {journal} {Annals of Physics}\ }\textbf {\bibinfo
  {volume} {417}},\ \bibinfo {pages} {168102} (\bibinfo {year} {2020})},\
  \bibinfo {note} {eliashberg theory at 60: Strong-coupling superconductivity
  and beyond}\BibitemShut {NoStop}%
\bibitem [{\citenamefont {Bauer}\ \emph {et~al.}(2012)\citenamefont {Bauer},
  \citenamefont {Han},\ and\ \citenamefont {Gunnarsson}}]{Bauer_2012}%
  \BibitemOpen
  \bibfield  {author} {\bibinfo {author} {\bibfnamefont {J.}~\bibnamefont
  {Bauer}}, \bibinfo {author} {\bibfnamefont {J.~E.}\ \bibnamefont {Han}},\
  and\ \bibinfo {author} {\bibfnamefont {O.}~\bibnamefont {Gunnarsson}},\
  }\href {https://doi.org/10.1088/0953-8984/24/49/492202} {\bibfield  {journal}
  {\bibinfo  {journal} {Journal of Physics: Condensed Matter}\ }\textbf
  {\bibinfo {volume} {24}},\ \bibinfo {pages} {492202} (\bibinfo {year}
  {2012})}\BibitemShut {NoStop}%
\bibitem [{\citenamefont {Bauer}\ \emph {et~al.}(2013)\citenamefont {Bauer},
  \citenamefont {Han},\ and\ \citenamefont {Gunnarsson}}]{PhysRevB.87.054507}%
  \BibitemOpen
  \bibfield  {author} {\bibinfo {author} {\bibfnamefont {J.}~\bibnamefont
  {Bauer}}, \bibinfo {author} {\bibfnamefont {J.~E.}\ \bibnamefont {Han}},\
  and\ \bibinfo {author} {\bibfnamefont {O.}~\bibnamefont {Gunnarsson}},\
  }\href {https://doi.org/10.1103/PhysRevB.87.054507} {\bibfield  {journal}
  {\bibinfo  {journal} {Phys. Rev. B}\ }\textbf {\bibinfo {volume} {87}},\
  \bibinfo {pages} {054507} (\bibinfo {year} {2013})}\BibitemShut {NoStop}%
\bibitem [{\citenamefont {Marsiglio}(1995)}]{MARSIGLIO199521}%
  \BibitemOpen
  \bibfield  {author} {\bibinfo {author} {\bibfnamefont {F.}~\bibnamefont
  {Marsiglio}},\ }\href
  {https://doi.org/https://doi.org/10.1016/0921-4534(95)00046-1} {\bibfield
  {journal} {\bibinfo  {journal} {Physica C: Superconductivity}\ }\textbf
  {\bibinfo {volume} {244}},\ \bibinfo {pages} {21} (\bibinfo {year}
  {1995})}\BibitemShut {NoStop}%
\bibitem [{\citenamefont {Marsiglio}(2022)}]{marsiglio2022impact}%
  \BibitemOpen
  \bibfield  {author} {\bibinfo {author} {\bibfnamefont {F.}~\bibnamefont
  {Marsiglio}},\ }\href@noop {} {\bibfield  {journal} {\bibinfo  {journal}
  {arXiv preprint arXiv:2205.10352}\ } (\bibinfo {year} {2022})}\BibitemShut
  {NoStop}%
\bibitem [{\citenamefont {Migdal}(1958)}]{migdal1958interaction}%
  \BibitemOpen
  \bibfield  {author} {\bibinfo {author} {\bibfnamefont {A.}~\bibnamefont
  {Migdal}},\ }\href@noop {} {\bibfield  {journal} {\bibinfo  {journal} {Sov.
  Phys. JETP}\ }\textbf {\bibinfo {volume} {7}},\ \bibinfo {pages} {996}
  (\bibinfo {year} {1958})}\BibitemShut {NoStop}%
\bibitem [{\citenamefont {Wu}\ \emph {et~al.}(2021)\citenamefont {Wu},
  \citenamefont {Zhang}, \citenamefont {Abanov},\ and\ \citenamefont
  {Chubukov}}]{PhysRevB.103.024522}%
  \BibitemOpen
  \bibfield  {author} {\bibinfo {author} {\bibfnamefont {Y.-M.}\ \bibnamefont
  {Wu}}, \bibinfo {author} {\bibfnamefont {S.-S.}\ \bibnamefont {Zhang}},
  \bibinfo {author} {\bibfnamefont {A.}~\bibnamefont {Abanov}},\ and\ \bibinfo
  {author} {\bibfnamefont {A.~V.}\ \bibnamefont {Chubukov}},\ }\href
  {https://doi.org/10.1103/PhysRevB.103.024522} {\bibfield  {journal} {\bibinfo
   {journal} {Phys. Rev. B}\ }\textbf {\bibinfo {volume} {103}},\ \bibinfo
  {pages} {024522} (\bibinfo {year} {2021})}\BibitemShut {NoStop}%
\bibitem [{\citenamefont {Christensen}\ and\ \citenamefont
  {Chubukov}(2021)}]{PhysRevB.104.L140501}%
  \BibitemOpen
  \bibfield  {author} {\bibinfo {author} {\bibfnamefont {M.~H.}\ \bibnamefont
  {Christensen}}\ and\ \bibinfo {author} {\bibfnamefont {A.~V.}\ \bibnamefont
  {Chubukov}},\ }\href {https://doi.org/10.1103/PhysRevB.104.L140501}
  {\bibfield  {journal} {\bibinfo  {journal} {Phys. Rev. B}\ }\textbf {\bibinfo
  {volume} {104}},\ \bibinfo {pages} {L140501} (\bibinfo {year}
  {2021})}\BibitemShut {NoStop}%
\bibitem [{\citenamefont {Brown}\ and\ \citenamefont
  {Churchill}(2009)}]{brown2009complex}%
  \BibitemOpen
  \bibfield  {author} {\bibinfo {author} {\bibfnamefont {J.~W.}\ \bibnamefont
  {Brown}}\ and\ \bibinfo {author} {\bibfnamefont {R.~V.}\ \bibnamefont
  {Churchill}},\ }\href@noop {} {\emph {\bibinfo {title} {Complex variables and
  applications}}}\ (\bibinfo  {publisher} {McGraw-Hill},\ \bibinfo {year}
  {2009})\BibitemShut {NoStop}%
\bibitem [{\citenamefont {Damascelli}(2004)}]{Damascelli_2004}%
  \BibitemOpen
  \bibfield  {author} {\bibinfo {author} {\bibfnamefont {A.}~\bibnamefont
  {Damascelli}},\ }\href {https://doi.org/10.1238/physica.topical.109a00061}
  {\bibfield  {journal} {\bibinfo  {journal} {Physica Scripta}\ }\textbf
  {\bibinfo {volume} {T109}},\ \bibinfo {pages} {61} (\bibinfo {year}
  {2004})}\BibitemShut {NoStop}%
\bibitem [{\citenamefont {Marsiglio}\ and\ \citenamefont
  {Carbotte}(2008)}]{Marsiglio:2008aa}%
  \BibitemOpen
  \bibfield  {author} {\bibinfo {author} {\bibfnamefont {F.}~\bibnamefont
  {Marsiglio}}\ and\ \bibinfo {author} {\bibfnamefont {J.~P.}\ \bibnamefont
  {Carbotte}},\ }\bibinfo {title} {Electron-phonon superconductivity}\
  (\bibinfo  {publisher} {Springer Berlin Heidelberg},\ \bibinfo {address}
  {Berlin, Heidelberg},\ \bibinfo {year} {2008})\ pp.\ \bibinfo {pages}
  {73--162}\BibitemShut {NoStop}%
\bibitem [{\citenamefont {Phan}\ and\ \citenamefont
  {Chubukov}(2022)}]{PhysRevB.105.064518}%
  \BibitemOpen
  \bibfield  {author} {\bibinfo {author} {\bibfnamefont {D.}~\bibnamefont
  {Phan}}\ and\ \bibinfo {author} {\bibfnamefont {A.~V.}\ \bibnamefont
  {Chubukov}},\ }\href {https://doi.org/10.1103/PhysRevB.105.064518} {\bibfield
   {journal} {\bibinfo  {journal} {Phys. Rev. B}\ }\textbf {\bibinfo {volume}
  {105}},\ \bibinfo {pages} {064518} (\bibinfo {year} {2022})}\BibitemShut
  {NoStop}%
\bibitem [{\citenamefont {Kusunose}\ \emph {et~al.}(2011)\citenamefont
  {Kusunose}, \citenamefont {Fuseya},\ and\ \citenamefont
  {Miyake}}]{doi:10.1143/JPSJ.80.044711}%
  \BibitemOpen
  \bibfield  {author} {\bibinfo {author} {\bibfnamefont {H.}~\bibnamefont
  {Kusunose}}, \bibinfo {author} {\bibfnamefont {Y.}~\bibnamefont {Fuseya}},\
  and\ \bibinfo {author} {\bibfnamefont {K.}~\bibnamefont {Miyake}},\
  }\href@noop {} {\bibfield  {journal} {\bibinfo  {journal} {Journal of the
  Physical Society of Japan}\ }\textbf {\bibinfo {volume} {80}},\ \bibinfo
  {pages} {044711} (\bibinfo {year} {2011})}\BibitemShut {NoStop}%
\bibitem [{\citenamefont {Pimenov}\ and\ \citenamefont
  {Chubukov}(2022{\natexlab{a}})}]{pimenov2021quantum}%
  \BibitemOpen
  \bibfield  {author} {\bibinfo {author} {\bibfnamefont {D.}~\bibnamefont
  {Pimenov}}\ and\ \bibinfo {author} {\bibfnamefont {A.~V.}\ \bibnamefont
  {Chubukov}},\ }\href {https://doi.org/10.1038/s41535-022-00457-3} {\bibfield
  {journal} {\bibinfo  {journal} {npj Quantum Materials}\ }\textbf {\bibinfo
  {volume} {7}},\ \bibinfo {pages} {45} (\bibinfo {year}
  {2022}{\natexlab{a}})}\BibitemShut {NoStop}%
\bibitem [{\citenamefont {Pimenov}\ and\ \citenamefont
  {Chubukov}(2022{\natexlab{b}})}]{pimenov2022odd}%
  \BibitemOpen
  \bibfield  {author} {\bibinfo {author} {\bibfnamefont {D.}~\bibnamefont
  {Pimenov}}\ and\ \bibinfo {author} {\bibfnamefont {A.~V.}\ \bibnamefont
  {Chubukov}},\ }\href@noop {} {\bibfield  {journal} {\bibinfo  {journal}
  {arXiv preprint arXiv:2206.01783}\ } (\bibinfo {year}
  {2022}{\natexlab{b}})}\BibitemShut {NoStop}%
\bibitem [{\citenamefont {Kosterlitz}(1974)}]{Kosterlitz_1974}%
  \BibitemOpen
  \bibfield  {author} {\bibinfo {author} {\bibfnamefont {J.~M.}\ \bibnamefont
  {Kosterlitz}},\ }\href {https://doi.org/10.1088/0022-3719/7/6/005} {\bibfield
   {journal} {\bibinfo  {journal} {Journal of Physics C: Solid State Physics}\
  }\textbf {\bibinfo {volume} {7}},\ \bibinfo {pages} {1046} (\bibinfo {year}
  {1974})}\BibitemShut {NoStop}%
\bibitem [{\citenamefont {Kaplan}\ \emph {et~al.}(2009)\citenamefont {Kaplan},
  \citenamefont {Lee}, \citenamefont {Son},\ and\ \citenamefont
  {Stephanov}}]{PhysRevD.80.125005}%
  \BibitemOpen
  \bibfield  {author} {\bibinfo {author} {\bibfnamefont {D.~B.}\ \bibnamefont
  {Kaplan}}, \bibinfo {author} {\bibfnamefont {J.-W.}\ \bibnamefont {Lee}},
  \bibinfo {author} {\bibfnamefont {D.~T.}\ \bibnamefont {Son}},\ and\ \bibinfo
  {author} {\bibfnamefont {M.~A.}\ \bibnamefont {Stephanov}},\ }\href
  {https://doi.org/10.1103/PhysRevD.80.125005} {\bibfield  {journal} {\bibinfo
  {journal} {Phys. Rev. D}\ }\textbf {\bibinfo {volume} {80}},\ \bibinfo
  {pages} {125005} (\bibinfo {year} {2009})}\BibitemShut {NoStop}%
\bibitem [{\citenamefont {Nelson}\ and\ \citenamefont
  {Kosterlitz}(1977)}]{PhysRevLett.39.1201}%
  \BibitemOpen
  \bibfield  {author} {\bibinfo {author} {\bibfnamefont {D.~R.}\ \bibnamefont
  {Nelson}}\ and\ \bibinfo {author} {\bibfnamefont {J.~M.}\ \bibnamefont
  {Kosterlitz}},\ }\href {https://doi.org/10.1103/PhysRevLett.39.1201}
  {\bibfield  {journal} {\bibinfo  {journal} {Phys. Rev. Lett.}\ }\textbf
  {\bibinfo {volume} {39}},\ \bibinfo {pages} {1201} (\bibinfo {year}
  {1977})}\BibitemShut {NoStop}%
\bibitem [{Note1()}]{Note1}%
  \BibitemOpen
  \bibinfo {note} {Note that our system has both time-reversal and
  translational invariance, which are necessary conditions for $n_s =1$ (in
  proper units) according to Leggett's theorem \cite
  {Leggetttheorem}.}\BibitemShut {Stop}%
\bibitem [{\citenamefont {Schmalian}(2022)}]{SchmalianComment}%
  \BibitemOpen
  \bibfield  {author} {\bibinfo {author} {\bibfnamefont {J.}~\bibnamefont
  {Schmalian}},\ }\href@noop {} {\bibfield  {journal} {\bibinfo  {journal}
  {Journal Club for Condensed Matter Physics}\ } (\bibinfo {year}
  {2022})}\BibitemShut {NoStop}%
\bibitem [{\citenamefont {Berezinskii}(1974)}]{berezinskii1974new}%
  \BibitemOpen
  \bibfield  {author} {\bibinfo {author} {\bibfnamefont {V.}~\bibnamefont
  {Berezinskii}},\ }\href@noop {} {\bibfield  {journal} {\bibinfo  {journal}
  {JETP Lett}\ }\textbf {\bibinfo {volume} {20}},\ \bibinfo {pages} {287}
  (\bibinfo {year} {1974})}\BibitemShut {NoStop}%
\bibitem [{\citenamefont {Emery}\ and\ \citenamefont
  {Kivelson}(1992)}]{Kivelson_1992}%
  \BibitemOpen
  \bibfield  {author} {\bibinfo {author} {\bibfnamefont {V.~J.}\ \bibnamefont
  {Emery}}\ and\ \bibinfo {author} {\bibfnamefont {S.}~\bibnamefont
  {Kivelson}},\ }\href {https://doi.org/10.1103/PhysRevB.46.10812} {\bibfield
  {journal} {\bibinfo  {journal} {Phys. Rev. B}\ }\textbf {\bibinfo {volume}
  {46}},\ \bibinfo {pages} {10812} (\bibinfo {year} {1992})}\BibitemShut
  {NoStop}%
\bibitem [{\citenamefont {Golubov}\ \emph {et~al.}(2009)\citenamefont
  {Golubov}, \citenamefont {Tanaka}, \citenamefont {Asano},\ and\ \citenamefont
  {Tanuma}}]{Golubov_2009}%
  \BibitemOpen
  \bibfield  {author} {\bibinfo {author} {\bibfnamefont {A.~A.}\ \bibnamefont
  {Golubov}}, \bibinfo {author} {\bibfnamefont {Y.}~\bibnamefont {Tanaka}},
  \bibinfo {author} {\bibfnamefont {Y.}~\bibnamefont {Asano}},\ and\ \bibinfo
  {author} {\bibfnamefont {Y.}~\bibnamefont {Tanuma}},\ }\href
  {https://doi.org/10.1088/0953-8984/21/16/164208} {\bibfield  {journal}
  {\bibinfo  {journal} {Journal of Physics: Condensed Matter}\ }\textbf
  {\bibinfo {volume} {21}},\ \bibinfo {pages} {164208} (\bibinfo {year}
  {2009})}\BibitemShut {NoStop}%
\bibitem [{\citenamefont {Tanaka}\ \emph {et~al.}(2012)\citenamefont {Tanaka},
  \citenamefont {Sato},\ and\ \citenamefont
  {Nagaosa}}]{doi:10.1143/JPSJ.81.011013}%
  \BibitemOpen
  \bibfield  {author} {\bibinfo {author} {\bibfnamefont {Y.}~\bibnamefont
  {Tanaka}}, \bibinfo {author} {\bibfnamefont {M.}~\bibnamefont {Sato}},\ and\
  \bibinfo {author} {\bibfnamefont {N.}~\bibnamefont {Nagaosa}},\ }\href@noop
  {} {\bibfield  {journal} {\bibinfo  {journal} {Journal of the Physical
  Society of Japan}\ }\textbf {\bibinfo {volume} {81}},\ \bibinfo {pages}
  {011013} (\bibinfo {year} {2012})}\BibitemShut {NoStop}%
\bibitem [{\citenamefont {Linder}\ and\ \citenamefont
  {Balatsky}(2019)}]{RevModPhys.91.045005}%
  \BibitemOpen
  \bibfield  {author} {\bibinfo {author} {\bibfnamefont {J.}~\bibnamefont
  {Linder}}\ and\ \bibinfo {author} {\bibfnamefont {A.~V.}\ \bibnamefont
  {Balatsky}},\ }\href {https://doi.org/10.1103/RevModPhys.91.045005}
  {\bibfield  {journal} {\bibinfo  {journal} {Rev. Mod. Phys.}\ }\textbf
  {\bibinfo {volume} {91}},\ \bibinfo {pages} {045005} (\bibinfo {year}
  {2019})}\BibitemShut {NoStop}%
\bibitem [{\citenamefont {Di~Bernardo}\ \emph
  {et~al.}(2015{\natexlab{a}})\citenamefont {Di~Bernardo}, \citenamefont
  {Diesch}, \citenamefont {Gu}, \citenamefont {Linder}, \citenamefont
  {Divitini}, \citenamefont {Ducati}, \citenamefont {Scheer}, \citenamefont
  {Blamire},\ and\ \citenamefont {Robinson}}]{DiBernardo2015a}%
  \BibitemOpen
  \bibfield  {author} {\bibinfo {author} {\bibfnamefont {A.}~\bibnamefont
  {Di~Bernardo}}, \bibinfo {author} {\bibfnamefont {S.}~\bibnamefont {Diesch}},
  \bibinfo {author} {\bibfnamefont {Y.}~\bibnamefont {Gu}}, \bibinfo {author}
  {\bibfnamefont {J.}~\bibnamefont {Linder}}, \bibinfo {author} {\bibfnamefont
  {G.}~\bibnamefont {Divitini}}, \bibinfo {author} {\bibfnamefont
  {C.}~\bibnamefont {Ducati}}, \bibinfo {author} {\bibfnamefont
  {E.}~\bibnamefont {Scheer}}, \bibinfo {author} {\bibfnamefont {M.~G.}\
  \bibnamefont {Blamire}},\ and\ \bibinfo {author} {\bibfnamefont {J.~W.~A.}\
  \bibnamefont {Robinson}},\ }\href {https://doi.org/10.1038/ncomms9053}
  {\bibfield  {journal} {\bibinfo  {journal} {Nature Communications}\ }\textbf
  {\bibinfo {volume} {6}},\ \bibinfo {pages} {8053} (\bibinfo {year}
  {2015}{\natexlab{a}})}\BibitemShut {NoStop}%
\bibitem [{\citenamefont {Di~Bernardo}\ \emph
  {et~al.}(2015{\natexlab{b}})\citenamefont {Di~Bernardo}, \citenamefont
  {Salman}, \citenamefont {Wang}, \citenamefont {Amado}, \citenamefont
  {Egilmez}, \citenamefont {Flokstra}, \citenamefont {Suter}, \citenamefont
  {Lee}, \citenamefont {Zhao}, \citenamefont {Prokscha}, \citenamefont
  {Morenzoni}, \citenamefont {Blamire}, \citenamefont {Linder},\ and\
  \citenamefont {Robinson}}]{PhysRevX.5.041021}%
  \BibitemOpen
  \bibfield  {author} {\bibinfo {author} {\bibfnamefont {A.}~\bibnamefont
  {Di~Bernardo}}, \bibinfo {author} {\bibfnamefont {Z.}~\bibnamefont {Salman}},
  \bibinfo {author} {\bibfnamefont {X.~L.}\ \bibnamefont {Wang}}, \bibinfo
  {author} {\bibfnamefont {M.}~\bibnamefont {Amado}}, \bibinfo {author}
  {\bibfnamefont {M.}~\bibnamefont {Egilmez}}, \bibinfo {author} {\bibfnamefont
  {M.~G.}\ \bibnamefont {Flokstra}}, \bibinfo {author} {\bibfnamefont
  {A.}~\bibnamefont {Suter}}, \bibinfo {author} {\bibfnamefont {S.~L.}\
  \bibnamefont {Lee}}, \bibinfo {author} {\bibfnamefont {J.~H.}\ \bibnamefont
  {Zhao}}, \bibinfo {author} {\bibfnamefont {T.}~\bibnamefont {Prokscha}},
  \bibinfo {author} {\bibfnamefont {E.}~\bibnamefont {Morenzoni}}, \bibinfo
  {author} {\bibfnamefont {M.~G.}\ \bibnamefont {Blamire}}, \bibinfo {author}
  {\bibfnamefont {J.}~\bibnamefont {Linder}},\ and\ \bibinfo {author}
  {\bibfnamefont {J.~W.~A.}\ \bibnamefont {Robinson}},\ }\href
  {https://doi.org/10.1103/PhysRevX.5.041021} {\bibfield  {journal} {\bibinfo
  {journal} {Phys. Rev. X}\ }\textbf {\bibinfo {volume} {5}},\ \bibinfo {pages}
  {041021} (\bibinfo {year} {2015}{\natexlab{b}})}\BibitemShut {NoStop}%
\bibitem [{\citenamefont {Balatsky}\ and\ \citenamefont
  {Abrahams}(1992)}]{PhysRevB.45.13125}%
  \BibitemOpen
  \bibfield  {author} {\bibinfo {author} {\bibfnamefont {A.}~\bibnamefont
  {Balatsky}}\ and\ \bibinfo {author} {\bibfnamefont {E.}~\bibnamefont
  {Abrahams}},\ }\href {https://doi.org/10.1103/PhysRevB.45.13125} {\bibfield
  {journal} {\bibinfo  {journal} {Phys. Rev. B}\ }\textbf {\bibinfo {volume}
  {45}},\ \bibinfo {pages} {13125} (\bibinfo {year} {1992})}\BibitemShut
  {NoStop}%
\bibitem [{\citenamefont {Sukhachov}\ \emph {et~al.}(2019)\citenamefont
  {Sukhachov}, \citenamefont {Juri\ifmmode \check{c}\else
  \v{c}\fi{}i\ifmmode~\acute{c}\else \'{c}\fi{}},\ and\ \citenamefont
  {Balatsky}}]{PhysRevB.100.180502}%
  \BibitemOpen
  \bibfield  {author} {\bibinfo {author} {\bibfnamefont {P.~O.}\ \bibnamefont
  {Sukhachov}}, \bibinfo {author} {\bibfnamefont {V.}~\bibnamefont
  {Juri\ifmmode \check{c}\else \v{c}\fi{}i\ifmmode~\acute{c}\else
  \'{c}\fi{}}},\ and\ \bibinfo {author} {\bibfnamefont {A.~V.}\ \bibnamefont
  {Balatsky}},\ }\href {https://doi.org/10.1103/PhysRevB.100.180502} {\bibfield
   {journal} {\bibinfo  {journal} {Phys. Rev. B}\ }\textbf {\bibinfo {volume}
  {100}},\ \bibinfo {pages} {180502} (\bibinfo {year} {2019})}\BibitemShut
  {NoStop}%
\bibitem [{\citenamefont {Abrahams}\ \emph {et~al.}(1993)\citenamefont
  {Abrahams}, \citenamefont {Balatsky}, \citenamefont {Schrieffer},\ and\
  \citenamefont {Allen}}]{PhysRevB.47.513}%
  \BibitemOpen
  \bibfield  {author} {\bibinfo {author} {\bibfnamefont {E.}~\bibnamefont
  {Abrahams}}, \bibinfo {author} {\bibfnamefont {A.}~\bibnamefont {Balatsky}},
  \bibinfo {author} {\bibfnamefont {J.~R.}\ \bibnamefont {Schrieffer}},\ and\
  \bibinfo {author} {\bibfnamefont {P.~B.}\ \bibnamefont {Allen}},\ }\href
  {https://doi.org/10.1103/PhysRevB.47.513} {\bibfield  {journal} {\bibinfo
  {journal} {Phys. Rev. B}\ }\textbf {\bibinfo {volume} {47}},\ \bibinfo
  {pages} {513} (\bibinfo {year} {1993})}\BibitemShut {NoStop}%
\bibitem [{\citenamefont {Langmann}\ \emph {et~al.}(2022)\citenamefont
  {Langmann}, \citenamefont {Hainzl}, \citenamefont {Seiringer},\ and\
  \citenamefont {Balatsky}}]{langmannOF}%
  \BibitemOpen
  \bibfield  {author} {\bibinfo {author} {\bibfnamefont {E.}~\bibnamefont
  {Langmann}}, \bibinfo {author} {\bibfnamefont {C.}~\bibnamefont {Hainzl}},
  \bibinfo {author} {\bibfnamefont {R.}~\bibnamefont {Seiringer}},\ and\
  \bibinfo {author} {\bibfnamefont {A.~V.}\ \bibnamefont {Balatsky}},\ }\href
  {https://arxiv.org/abs/2207.01825} {\bibfield  {journal} {\bibinfo  {journal}
  {arXiv preprint arXiv:2207.01825}\ } (\bibinfo {year} {2022})}\BibitemShut
  {NoStop}%
\bibitem [{\citenamefont {Schrodi}\ \emph {et~al.}(2021)\citenamefont
  {Schrodi}, \citenamefont {Aperis},\ and\ \citenamefont
  {Oppeneer}}]{PhysRevB.104.174518}%
  \BibitemOpen
  \bibfield  {author} {\bibinfo {author} {\bibfnamefont {F.}~\bibnamefont
  {Schrodi}}, \bibinfo {author} {\bibfnamefont {A.}~\bibnamefont {Aperis}},\
  and\ \bibinfo {author} {\bibfnamefont {P.~M.}\ \bibnamefont {Oppeneer}},\
  }\href {https://doi.org/10.1103/PhysRevB.104.174518} {\bibfield  {journal}
  {\bibinfo  {journal} {Phys. Rev. B}\ }\textbf {\bibinfo {volume} {104}},\
  \bibinfo {pages} {174518} (\bibinfo {year} {2021})}\BibitemShut {NoStop}%
\bibitem [{\citenamefont {Allen}\ and\ \citenamefont
  {Mitrovi{\'c}}(1983)}]{ALLEN19831}%
  \BibitemOpen
  \bibfield  {author} {\bibinfo {author} {\bibfnamefont {P.~B.}\ \bibnamefont
  {Allen}}\ and\ \bibinfo {author} {\bibfnamefont {B.}~\bibnamefont
  {Mitrovi{\'c}}}\ }(\bibinfo  {publisher} {Academic Press},\ \bibinfo {year}
  {1983})\ pp.\ \bibinfo {pages} {1--92}\BibitemShut {NoStop}%
\bibitem [{\citenamefont {Ghosh}\ \emph {et~al.}(2020)\citenamefont {Ghosh},
  \citenamefont {Smidman}, \citenamefont {Shang}, \citenamefont {Annett},
  \citenamefont {Hillier}, \citenamefont {Quintanilla},\ and\ \citenamefont
  {Yuan}}]{Ghosh_2020}%
  \BibitemOpen
  \bibfield  {author} {\bibinfo {author} {\bibfnamefont {S.~K.}\ \bibnamefont
  {Ghosh}}, \bibinfo {author} {\bibfnamefont {M.}~\bibnamefont {Smidman}},
  \bibinfo {author} {\bibfnamefont {T.}~\bibnamefont {Shang}}, \bibinfo
  {author} {\bibfnamefont {J.~F.}\ \bibnamefont {Annett}}, \bibinfo {author}
  {\bibfnamefont {A.~D.}\ \bibnamefont {Hillier}}, \bibinfo {author}
  {\bibfnamefont {J.}~\bibnamefont {Quintanilla}},\ and\ \bibinfo {author}
  {\bibfnamefont {H.}~\bibnamefont {Yuan}},\ }\href
  {https://doi.org/10.1088/1361-648x/abaa06} {\bibfield  {journal} {\bibinfo
  {journal} {Journal of Physics: Condensed Matter}\ }\textbf {\bibinfo {volume}
  {33}},\ \bibinfo {pages} {033001} (\bibinfo {year} {2020})}\BibitemShut
  {NoStop}%
\bibitem [{\citenamefont {Slagle}\ and\ \citenamefont
  {Fu}(2020)}]{PhysRevB.102.235423}%
  \BibitemOpen
  \bibfield  {author} {\bibinfo {author} {\bibfnamefont {K.}~\bibnamefont
  {Slagle}}\ and\ \bibinfo {author} {\bibfnamefont {L.}~\bibnamefont {Fu}},\
  }\href {https://doi.org/10.1103/PhysRevB.102.235423} {\bibfield  {journal}
  {\bibinfo  {journal} {Phys. Rev. B}\ }\textbf {\bibinfo {volume} {102}},\
  \bibinfo {pages} {235423} (\bibinfo {year} {2020})}\BibitemShut {NoStop}%
\bibitem [{\citenamefont {Cr{\'e}pel}\ and\ \citenamefont
  {Fu}(2021)}]{doi:10.1126/sciadv.abh2233}%
  \BibitemOpen
  \bibfield  {author} {\bibinfo {author} {\bibfnamefont {V.}~\bibnamefont
  {Cr{\'e}pel}}\ and\ \bibinfo {author} {\bibfnamefont {L.}~\bibnamefont
  {Fu}},\ }\href@noop {} {\bibfield  {journal} {\bibinfo  {journal} {Science
  Advances}\ }\textbf {\bibinfo {volume} {7}},\ \bibinfo {pages} {eabh2233}
  (\bibinfo {year} {2021})}\BibitemShut {NoStop}%
\bibitem [{\citenamefont {Cr\'epel}\ \emph {et~al.}(2022)\citenamefont
  {Cr\'epel}, \citenamefont {Cea}, \citenamefont {Fu},\ and\ \citenamefont
  {Guinea}}]{PhysRevB.105.094506}%
  \BibitemOpen
  \bibfield  {author} {\bibinfo {author} {\bibfnamefont {V.}~\bibnamefont
  {Cr\'epel}}, \bibinfo {author} {\bibfnamefont {T.}~\bibnamefont {Cea}},
  \bibinfo {author} {\bibfnamefont {L.}~\bibnamefont {Fu}},\ and\ \bibinfo
  {author} {\bibfnamefont {F.}~\bibnamefont {Guinea}},\ }\href
  {https://doi.org/10.1103/PhysRevB.105.094506} {\bibfield  {journal} {\bibinfo
   {journal} {Phys. Rev. B}\ }\textbf {\bibinfo {volume} {105}},\ \bibinfo
  {pages} {094506} (\bibinfo {year} {2022})}\BibitemShut {NoStop}%
\bibitem [{\citenamefont {Schlawin}\ \emph {et~al.}(2019)\citenamefont
  {Schlawin}, \citenamefont {Cavalleri},\ and\ \citenamefont
  {Jaksch}}]{PhysRevLett.122.133602}%
  \BibitemOpen
  \bibfield  {author} {\bibinfo {author} {\bibfnamefont {F.}~\bibnamefont
  {Schlawin}}, \bibinfo {author} {\bibfnamefont {A.}~\bibnamefont
  {Cavalleri}},\ and\ \bibinfo {author} {\bibfnamefont {D.}~\bibnamefont
  {Jaksch}},\ }\href {https://doi.org/10.1103/PhysRevLett.122.133602}
  {\bibfield  {journal} {\bibinfo  {journal} {Phys. Rev. Lett.}\ }\textbf
  {\bibinfo {volume} {122}},\ \bibinfo {pages} {133602} (\bibinfo {year}
  {2019})}\BibitemShut {NoStop}%
\bibitem [{\citenamefont {Abanov}\ and\ \citenamefont
  {Chubukov}(2020)}]{abanov}%
  \BibitemOpen
  \bibfield  {author} {\bibinfo {author} {\bibfnamefont {A.}~\bibnamefont
  {Abanov}}\ and\ \bibinfo {author} {\bibfnamefont {A.~V.}\ \bibnamefont
  {Chubukov}},\ }\href {https://doi.org/10.1103/PhysRevB.102.024524} {\bibfield
   {journal} {\bibinfo  {journal} {Phys. Rev. B}\ }\textbf {\bibinfo {volume}
  {102}},\ \bibinfo {pages} {024524} (\bibinfo {year} {2020})}\BibitemShut
  {NoStop}%
\bibitem [{\citenamefont {Classen}\ and\ \citenamefont
  {Chubukov}(2021)}]{laura}%
  \BibitemOpen
  \bibfield  {author} {\bibinfo {author} {\bibfnamefont {L.}~\bibnamefont
  {Classen}}\ and\ \bibinfo {author} {\bibfnamefont {A.}~\bibnamefont
  {Chubukov}},\ }\href {https://doi.org/10.1103/PhysRevB.104.125120} {\bibfield
   {journal} {\bibinfo  {journal} {Phys. Rev. B}\ }\textbf {\bibinfo {volume}
  {104}},\ \bibinfo {pages} {125120} (\bibinfo {year} {2021})}\BibitemShut
  {NoStop}%
\end{thebibliography}%
\bibliographystyle{apsrev4-2}

\end{document}